\documentclass[aps,prb,preprint]{revtex4}
\usepackage{subfigure}
\usepackage{graphicx}
\usepackage{dcolumn}
\usepackage{bm}
\usepackage{textcomp}
\usepackage{amsmath}
\usepackage{amssymb}
\usepackage{epstopdf}
\graphicspath{{./FIGS/}}

\makeatletter
\newenvironment{figurehere}
{\def\@captype{figure}}
{}
\makeatother
\begin{document}
\preprint{}

\title{Electronic and Thermoelectric Properties of Few-Layer Transition Metal Dichalcogenides}
\author{Darshana \surname{Wickramaratne}}
\affiliation{Department of Electrical Engineering,
University of California, Riverside, CA 92521-0204}

\author{Ferdows \surname{Zahid}}
\affiliation{Department of Physics and the Center of 
Theoretical and Computational Physics, 
The University of Hong Kong, Pokfulam Road, Hong Kong SAR, China}

\author{Roger K. \surname{Lake}}
\affiliation{Department of Electrical Engineering,
University of California, Riverside, CA 92521}

\makeatother

\begin{abstract}
The electronic and thermoelectric properties 
of one to four monolayers of MoS$_{2}$, MoSe$_{2}$, WS$_{2}$, 
and WSe$_{2}$ are calculated.  
%
%
%
For few layer thicknesses,
the near degeneracies of the conduction band $K$ and $\Sigma$ valleys
and the valence band $\Gamma$ and $K$ valleys
enhance the n-type and p-type thermoelectric performance.
The interlayer hybridization and energy level splitting 
determine how the number of modes within $k_BT$ of a valley minimum 
changes with layer thickness.
In all cases, the maximum ZT coincides with the greatest near-degeneracy within
$k_BT$ of the band edge that results in the sharpest turn-on of the density
of modes.
The thickness at which this maximum occurs is, in general, not a monolayer.
The transition from few layers to bulk is discussed.
Effective masses, energy gaps, power-factors, and ZT values are tabulated for all 
materials and layer thicknesses.
\end{abstract}
\maketitle

\section{INTRODUCTION}\label{sec:Introduction}
Semiconducting, transition-metal dichalcogenides (TMDCs)
exhibit promising
electronic \cite{MoS2:transistor:Kis,Sayeef:MoS2,Javey_WSe2,KAlam_MoS2FET_TED12,PYe_MoS2_FETs},
opto-electronic \cite{Galli:MoS2:JPC07} and spintronic \cite{MoS2:Yao:spin} properties.
Single monolayers (three atomic layers) can be either exfoliated
or grown with chemically stable surfaces.
The electronic, optical, and spin properties of
monolayers are qualitatively different from those of the bulk.
The band gap changes from indirect to direct, and the valence
band edges at the $K$ and $K'$ points become spin 
polarized.\cite{Galli:MoS2:JPC07,MoS2:Yao:spin}
These materials are discussed in a number of recent 
reviews.\cite{Novoselov_Neto_2D_rev12,Strano_2D_rev_NNano12,Geim_Grigorieva_review_Nat13,2D_materials_ChemRev13,OSU_2D_rev_ACSNano13} 

Experimental studies conducted on 
a different set of two-dimensional materials, namely
Bi$_{2}$Te$_{3}$ and Bi$_{2}$Se$_{3}$,
demonstrated an improvement in their thermoelectric
performance as their thickness was reduced. 
\cite{Bi2Se3:JAC:Xie, goyal_Balandin}
A large increase in ZT has been theoretically predicted
for monolayer Bi$_{2}$Te$_{3}$ compared to that of the
bulk.\cite{Moore_TI_thermo_PRL10,Zahid_Lake,Lundstrom_Jesse_Bi2Te3}  
This enhancement in ZT results from the unique, step-function shape  
of the density of modes at the
valence band edge of a single quintuple layer.\cite{Zahid_Lake,Lundstrom_Jesse_Bi2Te3} 
The shape of the density of modes increases the 
power factor, and the increase in the power factor increases ZT.
For Bi$_{2}$Te$_{3}$,
the large enhancement in the power factor and 
in ZT only occurs for a monolayer.
For bilayer and trilayer Bi$_{2}$Te$_{3}$, the step-like shape of the
density of modes disappears, and
the calculated values of ZT are either
slightly higher\cite{Zahid_Lake_unpub} 
or slightly lower\cite{Lundstrom_Jesse_Bi2Te3} than that of the bulk.

Prior experimental and theoretical investigations of the thermoelectric
performance of transition metal dichalcogenides have focused on either
bulk or monolayer materials.
\cite{thermoelectric_power_TiS2,Imai_thermopower_TiS2,pressure_MoS2_Zhang_JAP_2013, Galli:MoS2:JPC07, photothermoelectric:MoS2,Liang_MX2_thermoelectric,Whangbo_MX2_thermoelectric}.
There has not been a study of the effect of film thickness on the power
factor and ZT in the transition metal dichalcogenides.
It is not known whether the power factor and ZT are maximum
at monolayer thickness or at some other thickness.

This work theoretically investigates the electronic
properties and the thermoelectric performance 
of bulk and one to four monolayers of 4 different TMDC materials: 
MoS$_2$, MoSe$_2$, WS$_2$, and WSe$_2$.
The goal is to understand how their electronic and thermoelectric properties
vary with thickness.
%
%
%
Similar to monolayer Bi$_{2}$Te$_{3}$, 
the increase in ZT for the ultrathin films
results from an enhanced degeneracy or near-degeneracy
of the band edges.
In the TMDCs, at few layer thicknesses, different
valleys become nearly degenerate with energy
differences of less than $k_BT$ at room temperature.
Because of 
weak interlayer coupling at certain valleys,
additional bands from additional layers lie within
$k_BT$ of the bandedges for few layer thicknesses.
The increased degeneracy results in a sharper turn on of the
the density of modes near the band edges.
In all cases, 
the thickness with the sharpest increase in the density of modes
has the largest value for ZT.
For the semiconducting TMDCs considered here, that optimum
thickness is not, in general, a single monolayer.
%

\section{THEORETICAL METHODS}\label{sec:Method}
\textit{Ab-initio} calculations
of the bulk and few-layer structures (one to four layers) are carried out 
using density functional theory (DFT)
with a projector augmented wave method \cite{PAW} and the 
Perdew-Burke-Ernzerhof (PBE) type generalized gradient 
approximation \cite{perdew:1996:PBE,ernzerhof:1999:PBE_test} 
as implemented in the Vienna {\em ab-initio} simulation package (VASP). 
\cite{VASP1,VASP2}
The vdW interactions in MoSe$_{2}$ and MoS$_{2}$ are accounted for using a 
semi-empirical correction to the Kohn-Sham energies when optimizing the bulk
structures (optimization of WS$_{2}$ and WSe$_{2}$ structures are done
at the PBE level since the semi-empirical parameters for tungsten are currently not 
described by the dispersion potential).\cite{Grimme_DFT_D2}
The Monkhorst-Pack scheme is used for the integration of the Brillouin zone
with a k-mesh of 12 x 12 x 6 for the bulk structures and 12 x 12 x 1 for the
thin-films.
The energy cutoff of the plane wave basis is 300 eV.
All of the the electronic bandstructure calculations
include spin-orbit coupling.
Calculations are also performed without spin-orbit coupling and the 
results are compared.

To verify the results of the PBE calculations, 
the electronic structure of 1L, 2L, 3L and 4L MoS$_{2}$ 
are calculated using the much more computationally expensive hybrid
Heyd-Scuseria-Ernzerhof (HSE) functional.\cite{HSE_VASP}
The HSE calculations incorporate 25$\%$ short-range Hartree-Fock exchange.
The screening parameter $\mu$ is set to 0.4 $\AA^{-1}$.

The thermoelectric parameters are calculated from a
Landauer formalism using the \emph{ab-initio} derived
density of modes.\cite{Zahid_Lake,Lundstrom_Jesse_Bi2Te3,Liang_MX2_thermoelectric}
In the linear response regime, the
electronic conductivity ($\sigma$), the electronic thermal conductivity ($\kappa_{e}$),
and the Seebeck coefficient (S) are expressed as \cite{Lundstrom:TE:TB:JAP:2010,Klimeck_DOM_thermoelectric_JCE}
\begin{eqnarray}
\sigma &=& (2q^{2}/h)I_{0}\quad (\mathrm{\Omega^{-1} m^{-1}}),
\label{eq:sigma}\\
\kappa_{e} &=& (2Tk_{B}^{2}/h)(I_{2} - I_{1}^{2}/I_{0}) \quad (\mathrm{W  m^{-1} K^{-1}}),
\label{eq:Ke}\\
S &=& -(k_{B}/q)\frac{I_{1}}{I_{0}}\quad (\mathrm{V/K}),
\label{eq:S}\\
\mathrm{with} \nonumber \\
I_{j} &=& \frac{1}{L} \int_{-\infty}^{\infty} \left(\frac{E-E_{F}}{k_{B}T}\right)^{j} \bar{T}(E)\left(-\frac{\partial f_{0}}{\partial E}\right)dE  
\label{eq:Ij}
\end{eqnarray}
where $L$ is the device length, $q$ is the magnitude of the electron charge, $h$
is Planck's constant, and $k_B$ is Boltzmann's constant. 
The transmission function is
\begin{equation}
\bar{T}(E) = T(E)M(E)
\label{eq:TE}
\end{equation}
where M(E) as the density of modes (DOM). 
In the diffusive limit,
\begin{equation}
T(E)=\lambda(E)/L , 
\label{eq:TE_diff}
\end{equation}
and $\lambda(E)$ is the electron mean free path. 
When phonon scattering is dominant, 
the mean free path can be written as a constant, $\lambda(E)=\lambda_{0}$.
As discussed in Ref. [\onlinecite{Lundstrom:TE:analytic:JAP:2009}],
the transport distribution, $\Xi(E)$, arising from the
Boltzmann transport equation is related to the above quantities by
$\Xi(E) = \frac{2}{h} T(E) M(E)$.

The density of modes M(E)
can be defined as \cite{Lundstrom_Jesse_Bi2Te3, Lundstrom:TE:TB:JAP:2010}
\begin{equation}
M(E)=\left (\frac{L_{\perp}}{2\pi}\right)^{d-1} \int_{BZ}\sum_{k_{\perp}} \Theta(E-\epsilon({k_{\perp}}))dk_{\perp}^{d-1}
\label{eq.ME}
\end{equation}
where $d$ is the dimensionality of the system, $L_{\perp}$ are the dimensions of the structure perpendicular 
to the direction of transport ($L_{\perp}^{2}$ = W x t for d = 3, $L_{\perp}$ = W for d = 2; W = width
of the structure, t = film thickness), 
$\Theta$ is the unit step function, 
and $k_{\perp}$ refers to the $k$ states in the
first Brillouin zone perpendicular to the transport direction. 
Using Eq. (\ref{eq.ME}), M(E) of any material in any dimension can be
numerically evaluated from a given electronic band structure by counting the bands
that cross the energy of interest. 
The density of modes calculations are performed by 
integrating over the first Brillouin zone using a 
converged k point grid (51 x 51 x 10 k points for the
bulk structures and 51 x 51 x 1 for the thin films). 

We account for carrier scattering within each structure by 
fitting our calculated bulk electrical conductivity with bulk experimental
data.
An electron mean free path of $\lambda_{0}$ = 14 nm gives the best
agreement with experimental data on the Seebeck response of bulk
MoS$_{2}$ as a function of the electrical conductivity.\cite{Mansfield_MoS2, Dutta_MoS2_thermoelectric}
The bulk p-type electrical conductivity of MoS$_{2}$ at room temperature was reported
to be 5.1 $\Omega^{-1}$cm$^{-1}$  with a Seebeck coefficient of $\sim$450 $\mu$VK$^{-1}$ 
at a carrier concentration of 10$^{16}$ cm$^{-3}$.\cite{Dutta_MoS2_thermoelectric}
Using $\lambda_{0}$ = 14nm we obtain an electrical conductivity of 
4.97 $\Omega^{-1}$cm$^{-1}$  with a Seebeck coefficient of $\sim$398 $\mu$VK$^{-1}$ 
at the same carrier concentration.
This value of the mean free path is also consistent with a 
theoretically derived energy independent acoustic phonon-limited mean
free path ($\lambda_{0}$ = 14 nm) for electrons in monolayer 
MoS$_{2}$,\cite{MoS2_Thygesen_mfp} and was successfully
used to simulate and compare to experimental results of the transfer
characteristics of single layer MoS$_{2}$ field effect transistor.\cite{Sayeef:MoS2}
As an initial approximation of carrier scattering we use the same
$\lambda_{0}$ value to model the thermoelectric properties of all the
TMDC materials investigated in this study.

For the in-plane lattice thermal conductivity, a $\kappa_{l}$ value of 
19.5 $Wm^{-1}K^{-1}$ obtained from a molecular dynamics simulation
on {monolayer MoS$_{2}$ is used.\cite{Varshney_MoS2}
Prior experimental\cite{Chiritescu:WSe2} and theoretical\cite{Liang_MX2_thermoelectric}
studies of the lattice thermal conductivity in the TMDC materials have 
demonstrated that $\kappa_{L}$ does not vary significantly for the different TMDC compounds
studied here.
With the above quantities in hand, the power factor, $S^2 \sigma$, and the thermoelectric
figure of merit $ZT = S^2 \sigma T / (\kappa_l + \kappa_e)$ are determined. 

\section{RESULTS}\label{sec:Results}

All of the thermoelectric parameters are derived
from the calculated electronic bandstructures. 
Therefore, we begin this section with a discussion of the
calculated bandstructures.
The bandstructure calculations produce considerably 
more information than is required for calculating the 
thermoelectric parameters.
To preserve that information and contribute towards
a database of material parameters, extracted properties such as
effective masses and energy gaps at high symmetry points
are tabulated.
Figure \ref{Ek} shows the \emph{ab-initio} band structure of
one-layer (1L) through four-layer (4L) 
and bulk WS$_{2}$.
The large valence band splitting at the $K$-point and the direct-indirect gap
transition as the film thickness increases above 1L   
are features that occur in the other TMDC materials included as part of this study.
The last panel in Fig. \ref{Ek} illustrates the effect of decreasing layer thickness
on the bandgap for all of the materials studied.
The optimized lattice parameters of the bulk TMDC compounds are listed in Table \ref{tab:params}.
The results in Table \ref{tab:params} and Figure \ref{Ek} are consistent
with prior experimental characterization 
\cite{MoS2_MoSe2_structure_Jellinek,WS2_WSe2_structure_Jellinek, Kam_bulk_Egap_JPhysChem} 
and theoretical calculations of the bulk\cite{Jiang_GW_MoX2_WX2, deGroot_bulk_MX2}
and thin film\cite{Ding_1L_MX2,Kuc_quantum_confinement_TS2} crystal 
structures and electronic band structures.
The results of these electronic structure calculations at the high symmetry points
are summarized in Tables \ref{tab:Eff_Mass} and Table \ref{tab:Energy_gaps}.
Table \ref{tab:Eff_Mass} gives the relative effective masses, and
Table \ref{tab:Energy_gaps} gives the energy gaps.

A number of prior theoretical studies of the electronic structure of 
monolayer and few-layer TMDCs 
did not include spin-orbit interaction.
\cite{Guo_MX2_IEEE11, Heine_TS2_PRB11, Tang_MX2_PhysicaB11}
As a result, the band bandgaps reported in those studies are slightly larger.
For example the bandgaps reported in a prior PBE level calculation \cite{Tang_MX2_PhysicaB11}
are greater by 70 meV, 260 meV and and 284 meV for MoS$_{2}$ and MoSe$_{2}$, WS$_{2}$ 
and WSe$_{2}$ respectively when compared to our calculation results.
Without the inclusion of spin-orbit interaction, our values
for the bandgap of the monolayer TMDCs are consistent with the bandgaps
reported in these studies.
Including spin-orbit coupling results in a splitting 
of the valence bands, $\Delta_{SO}$, at $K$.
The spin orbit interaction shifts up one of the degenerate valence bands,
and this reduces the bandgap.
The degree of the energy shift
ranges from 39.6 meV for MoS$_{2}$ to 210.9 meV for WS$_{2}$.
The second degenerate valence band is shifted down by an energy 
that is also unique to each TMDC material; 
this ranges from 110.4 meV for MoS$_{2}$ to 316.2 meV for WSe$_{2}$.
For example the calculated $\Delta_{SO}$ energies of the monolayer TMDCs 
are 150 meV, 181 meV, 425 meV and 461 meV for MoS$_{2}$, MoSe$_{2}$, WS$_{2}$ and WSe$_{2}$,
 respectively.
This is in good agreement with a prior PBE level calculation \cite{Ramasub_MX2_PRB12}
that accounted for spin-orbit interaction which obtained $\Delta_{SO}$ values of 
146 meV, 183 meV, 425 meV and 461 meV for MoS$_{2}$, MoSe$_{2}$, WS$_{2}$ and WSe$_{2}$,  
respectively,
and a $\Delta_{SO}$ energy of 188 meV obtained for monolayer MoS$_{2}$ with the use of 
optical absorption experiments. \cite{MoS2_Mak_Heinz}

More sophisticated many-body \emph{ab-initio} calculations which include HSE 
or GW calculations have been reported in prior studies of the band structure
of monolayer \cite{Lambrecht_MX2_PRB12, Ramasub_MX2_PRB12, Yakobsen_MX2_PRB13,FZahid_MoS2}
and bilayer \cite{Lambrecht_MX2_PRB12,FZahid_MoS2}
structures of the molybdenum and tungsten dichalcogenides.
The values for $\Delta_{SO}$ resulting from these theories are only slightly changed
from those of the PBE model.
The $\Delta_{SO}$ values reported for monolayer
MoS$_{2}$, MoSe$_{2}$, WS$_{2}$ and WSe$_{2}$
with a GW (HSE) calculation are 164 (193) meV, 212 (261) meV, 456 (521) meV
and 501 (586) meV.\cite{Ramasub_MX2_PRB12}
The primary difference between the PBE and the HSE and GW calculations 
is an increase in the bandgap.
However, the PBE bandgap is large enough compared to the temperatures considered
that the exact magnitude of the bandgap has no effect on the thermoelectric
parameters.
An explicit comparison of the electronic structure and the thermoelectric parameters
calculated from the PBE and the HSE functionals for 1L - 4L MoS$_2$ 
is given below. 
%

Calculation of the thermoelectric parameters requires the
density of modes extracted from the electronic bandstructure
using Eq. (\ref{eq.ME}).
Figure \ref{DOM} shows the density of modes versus energy for
bulk, 1L, 2L, 3L, and 4L MoS$_{2}$, MoSe$_{2}$, WS$_{2}$, and WSe$_{2}$.
To compare the density of modes of the 
bulk structure with the thin-film  structures,
we divide the density of modes of the thin-film structures 
by their respective thickness, $t$.
As will be shown, for these TMDCs, 
small variations in the shape of the density of modes near the band edges can enhance
the power factor and subsequently ZT.
The thermoelectric properties of the bulk and thin-film
structures are calculated from Eqs. (\ref{eq:sigma}) - (\ref{eq:TE_diff})
using the density of modes shown in Fig. \ref{DOM}.

%

%
The Seebeck coefficient, electrical conductivity, power-factor (PF), and the
thermoelectric figure-of-merit (ZT) as a function of the reduced 
Fermi level, $\eta_{F}$ are shown in Figures \ref{Seebeck}, \ref{El_Conductivity},
\ref{PF}, and \ref{ZT}, respectively.
The reduced Fermi-level is $\eta_{F} = \frac{E_{F}-E_{C1}}{kT}$ for 
electrons in the conduction band, and 
$\eta_{F} = \frac{E_{F}-E_{V1}}{kT}$ for holes in the valence band.
$E_{C1}$ and $E_{V1}$ are the energies of the conduction and valence band
edges, respectively.
For each material and each thickness the maximum power factor and ZT occurs
for the conduction band states.
The peak conduction band and valence band
power factor and ZT for each 
structure and material at 77K, 150K and 300K
are summarized in Table \ref{tab:PF_max} and Table \ref{tab:ZT_max}, respectively.
For all materials, the few layer structures show a large increase in the 
values of their power factor and ZT 
compared to those of the bulk.

The peak n-type ZT values (and corresponding layer thicknesses)
for MoS$_{2}$, MoSe$_{2}$, WS$_{2}$ and WSe$_{2}$ are
2.23 (t=3L), 2.39 (t=2L), 2.03 (t=3L) and 1.91 (t=2L)
which is an improvement by a factor of 6.4, 8.2, 7.2 and 7.5 
over the respective bulk values.
These peak ZT values occur when the Fermi level is moved by
1.39kT, 1.55kT, 1.08kT and 1.39kT, respectively, below
the conduction band at T=300K.
This corresponds to electron carrier densities of
6.26 $\times$ 10$^{19}$ cm$^{-3}$, 5.74 $\times$ 10$^{19}$ cm$^{-3}$,
5.34 $\times$ 10$^{19}$ cm$^{-3}$ and 4.72 $\times$ 10$^{19}$ cm$^{-3}$
for MoS$_{2}$, MoSe$_{2}$, WS$_{2}$ and WSe$_{2}$
respectively.
The peak p-type ZT values (and corresponding layer thicknesses)
for MoS$_{2}$, MoSe$_{2}$, WS$_{2}$ and WSe$_{2}$ are
1.15 (t=2L), 0.81 (t=2L-4L), 0.76 (t=2L-3L) and 0.62 (t=1L-4L)
which is an improvement by a factor of 14.4, 10.1, 9.5 and 5.2 
over the respective bulk values.
These peak ZT values occur when the Fermi level is moved by
1.16kT, 1.01kT, 0.93kT and 0.85kT, respectively, above
the valence band at T=300K.
This corresponds to hole carrier densities of
7.12 $\times$ 10$^{19}$ cm$^{-3}$, 5.84 $\times$ 10$^{19}$ cm$^{-3}$,
4.02 $\times$ 10$^{19}$ cm$^{-3}$ and 3.91 $\times$ 10$^{19}$ cm$^{-3}$
for MoS$_{2}$, MoSe$_{2}$, WS$_{2}$ and WSe$_{2}$
respectively.
Of the four TMDC materials studied, MoS$_{2}$ is the only material
to exhibit a p-type and n-type ZT $>$ 1.
In contrast to Bi$_2$Te$_3$, the peak value of ZT does not 
occur in any of the materials at a monolayer thickness.

The Seebeck coefficients at the maximum n-type (p-type) ZT for
each material are 275 (245.6) $\mu$VK$^{-1}$, 287 (230.7) $\mu$VK$^{-1}$, 
279 (230.1) $\mu$VK$^{-1}$ and 276 (216.7) $\mu$VK$^{-1}$ for 
MoS$_{2}$, MoSe$_{2}$, WS$_{2}$ and WSe$_{2}$
respectively.
However, the Seebeck coefficients at the maximum
n-type (p-type) power factor for
each material are 167 (90.4) $\mu$VK$^{-1}$, 100 (185.8) $\mu$VK$^{-1}$, 
165 (177.1) $\mu$VK$^{-1}$ and 171 (172.1) $\mu$VK$^{-1}$ for 
MoS$_{2}$, MoSe$_{2}$, WS$_{2}$ and WSe$_{2}$, respectively. 
This is generally consistent with the conclusion
of a report on engineering the Seebeck coefficient
to obtain the maximum thermoelectric power factor.
\cite{optimal_Seebeck_Bandaru_APL09}

Without the inclusion of spin-orbit interaction
our values of the ballistic ZT for the monolayer TMDC materials
are consistent with a prior report on the monolayer thermoelectric properties 
of these TMDC materials.\cite{Liang_MX2_thermoelectric}
Our calculations show that without the inclusion of spin-orbit interaction
the peak n-type ZT values for all materials continue to occur at thicknesses
above a single monolayer.
The peak n-type ZT values (and corresponding layer thicknesses) without
spin-orbit interaction for MoS$_{2}$, MoSe$_{2}$, WS$_{2}$ and WSe$_{2}$ are
1.38 (t=3L), 1.52 (t=2L), 1.13 (t=4L) and 1.28 (t=2L).
However, the peak p-type ZT values without spin-orbit interaction 
occurs for a single monolayer for each TMDC material.
The p-type ZT values without spin-orbit interaction for
MoS$_{2}$, MoSe$_{2}$, WS$_{2}$ and WSe$_{2}$ are
1.42, 0.84, 0.90 and 0.69.

Recent electronic structure calculations using the Heyd-Scuseria-Ernzerhof (HSE)
hybrid functional \cite{Ferdows_MoS2} give
a bandgap that more accurately matches known experimental 
values.\cite{Ferdows_MoS2} 
To assess whether the trends in the thermoelectric parameters
predicted with the PBE functional 
are the same as those resulting from the HSE functional, 
we calculate the electronic band 
structure of 1L, 2L, 3L and 4L MoS$_{2}$
with both the PBE and the HSE functional 
and plot the results in
Fig. \ref{MoS2_PBE_HSE}.
Near the band edges, the 
HSE energies appear to be shifted with respect to the PBE energies.
The effective masses for the HSE band structures are lower by up to 17$\%$ 
for the conduction band valleys at $K$ and $\Sigma$ and are lower by up to 11$\%$ 
for the valence band valleys at $K$ and $\Gamma$.

To verify that the HSE functional leaves the thermoelectric trends
predicted from the PBE functional unchanged,
we compute the density-of-modes and thermoelectric performance 
of 1L, 2L and 3L MoS$_{2}$ using the HSE functional with the inclusion
of spin orbit coupling.
Figure \ref{MoS2_HSE} illustrates the DOM, Seebeck coefficient,  power factor and 
ZT for the 1L, 2L and 3L structures of MoS$_{2}$ computed with the HSE
functional.
The quantitative values do differ.
For the MoS$_2$ trilayer structure, the HSE (PBE) functionals give 
a peak n-type power factor of 0.41 (0.28) WK$^{-2}$m$^{-2}$ and 
a peak n-type ZT of 2.4 (2.2). 
However,
the HSE results for few-layer MoS$_{2}$ structures demonstrate 
the same trends in the shape of the density of modes
and the same trends in the values for the power factors 
and ZT. 
Both the HSE and PBE calculations show that
the turn-on of the density of modes is sharpest for the tri-layer 
structure resulting in maximum values for the power factor and ZT.  
Since the primary effect on the low energy states of the exact exchange in the HSE
functional is to shift the band edges with respect to those of a PBE calculation,
the trends resulting from the shape of the density of modes should be preserved.

Figure \ref{ZT_PF_dimension} summarizes the values from the PBE calculations for the 
peak n-type and p-type ZT and power factors
for each TMDC material and layer thickness.
In the n-type MoSe$_{2}$, WS$_{2}$ and WSe$_{2}$ structures, the peak power-factor and 
the peak ZT do not
occur at the same film thickness.
For example, in MoSe$_{2}$, 
a single monolayer has the highest power factor,  
and a bilayer has the highest ZT.
This can be explained by 
the increase in the  electronic thermal conductivity, $\kappa_{e}$
as the Fermi level is moved into the conduction band.

Figure \ref{kappa_eff} shows the ratio of the total thermal conductivity, 
$\kappa_{\rm tot}$, with respect to the lattice thermal conductivity, $\kappa_l$,
for each TMDC material.  
The two guide lines on each figure illustrate the reduced Fermi level position at 
which the maximum n-type power factor and ZT occurs. 
The ratio $\kappa_{\rm tot} / \kappa_l = 1 + \kappa_e / \kappa_l$ 
is higher at the Fermi level position where the
the maximum power factor occurs.
This increase in $\kappa_e$ explains why the peak power factor
and ZT occur at different Fermi energies and film thicknesses.

A number of recent studies report on the 
theoretical \cite{MoS2_kappaL_Mingo, MoS2_kappaL_1L_Zheng} 
and experimental values \cite{crossplane_kappaL_TMD_Muratore, MoS2_kappaL_Kis}
of the lattice thermal conductivity
on monolayer and few-layer TMDC materials with values of $\kappa_{l}$
ranging from 19 Wm$^{-1}$K$^{-1}$ to 83 Wm$^{-1}$K$^{-1}$.
Experimental measurements of the in-plane $\kappa_{l}$ 
in suspended samples of
MoS$_{2}$ \cite{MoS2_kappaL_Kis} find a value of 34.5 Wm$^{-1}$K$^{-1}$
for 1L MoS$_{2}$ and 52 Wm$^{-1}$K$^{-1}$ for few-layer MoS$_{2}$.
To assess whether the inequivalent $\kappa_{l}$ values for the monolayer and few-layer
TMDC films leave the predicted thermoelectric trends unchanged, 
we computed the thermoelectric 
parameters using $\kappa_{l}$=34.5 Wm$^{-1}$K$^{-1}$ 
for the monolayer and $\kappa_{l}$=52 Wm$^{-1}$K$^{-1}$ for the few-layer
TMDC films of each material.
The values of ZT differ compared to using $\kappa_{l}$=19 Wm$^{-1}$K$^{-1}$ 
for each film thickness.
For MoS$_{2}$, the room temperature n-type
ZT values using the thickness dependent (constant) $\kappa_{l}$ for the 
1L, 2L, 3L and 4L structures are 0.87 (1.35), 0.63 (1.15), 1.11 (2.23), 0.89 (1.78).
The maximum n-type ZT still occurs for the 3L structure and the minimum n-type
ZT still occurs for the 1L structure. 
The trends for all of the n-type materials are preserved when
a thickness dependent thermal conductivity is used.
All of the values are shown in Fig. \ref{ZT_kappa_thickness}(b).
For the n-type materials, changes in the density of modes are the dominant factor
determining the trends.
For p-type MoSe$_{2}$, WS$_{2}$, WSe$_{2}$, 
ZT varies little for different layer thicknesses when using a constant $\kappa_l$
as shown in Fig. \ref{ZT_PF_dimension}(a).
For p-type MoS$_{2}$, the difference between the maximum ZT of a bilayer
and the second highest value of a monolayer is small. 
Therefore, reducing the value of $\kappa_l$ for a monolayer from
52 to 35.4 WM$^{-1}$K$^{-1}$ is sufficient to cause the peak
value of ZT to occur at monolayer thickness for all 4 p-type materials
as shown in Fig. \ref{ZT_kappa_thickness}(a).

In an attempt to study the transition of the thermoelectric performance from
few-layer films to bulk like performance, we calculate the thermoelectric parameters
for an 8L film of WS$_{2}$.
Figure \ref{WS2_DOM_ZT_bulklike} illustrates the density of modes and the ZT for bulk, 3L and
8L WS$_{2}$.
The n-type 0.974 ZT value of the 8L film is a factor of 1.9 smaller than
that of the 4L film, but it is still a factor of 3.5 larger than that of the bulk. 
The p-type 0.163 ZT value of the 8L film is a factor of 4.7 smaller than
that of the 4L film, and it is a factor of 2.0 larger that that of the bulk.
Even at 8 monolayers, there is still an enhancement of the ZT value
compared to that of the bulk, and the enhancement is larger in the
n-type material.

The thermoelectric performance in the low dimensional structures is
enhanced by the more abrupt step-like shape of the
density of modes distribution.\cite{Lundstrom:TE:analytic:JAP:2009}
It is clear from Eq. (\ref{eq:Ij}), that with $E_F \leq 0$,
a step-function density
of modes removes all negative contributions to the integrand
of $I_1$ giving a maximum value for $I_1$.
The conduction band DOM distribution for the maximum and minimum
ZT structures for each material are plotted in Figure \ref{DOM_CB}.
In all cases, the DOM with the sharper turn-on at the band edge
gives rise the the maximum value for ZT.

\section{DISCUSSION}\label{sec:Discussion}

The enhancement in the thermoelectric performance of {\em few} monolayer
TMDC materials is in contrast to the enhanced thermoelectric performance
observed for only a single quintuple (QL) layer of p-type Bi$_{2}$Te$_{3}$.
Above 1 QL of Bi$_{2}$Te$_{3}$, the thermoelectric figure of merit approaches
the bulk ZT.  \cite{Zahid_Lake,Lundstrom_Jesse_Bi2Te3}.
The enhancement of ZT in n-type monolayer Bi$_{2}$Te$_{3}$ is minimal.
This difference in the effect of layer thickness on ZT in the two different classes
of materials can be explained by differences in the effect 
of thickness on the band-edge degeneracy and the density of modes.
The valence band of monolayer
Bi$_{2}$Te$_{3}$ is a ring in k-space that covers much of the Brillouin zone
as shown in Fig. 4(d) of Ref. \onlinecite{Lundstrom_Jesse_Bi2Te3}.
Thus, the integration over $k_{\perp}$ in Eq. (\ref{eq.ME}) jumps from zero in the
band gap to a finite number at the band edge resulting in a step-function 
turn-on of the valence band density of modes as seen in Fig. 3 of Ref.
\onlinecite{Zahid_Lake} and Fig. 2 of Ref. \onlinecite{Lundstrom_Jesse_Bi2Te3}.
The size of the ring in k-space quickly collapses for thicknesses above a monolayer,
and the large enhancement in ZT dissappears. 
In a parabolic band, the band edge is a point in k-space, and, in two-dimensions,
the density of modes turns on smoothly as $\sqrt{E}$.\cite{Lundstrom:TE:analytic:JAP:2009} 
The band edge of n-type monolayer Bi$_{2}$Te$_{3}$ remains parabolic resulting in
a smooth turn-on of the density-of-modes and no significant enhancement of ZT.

The bands of the TMDC materials also remain parabolic at the band edges, however the
conduction bands at the $K_c$ and the $\Sigma_c$ valleys become nearly degenerate
for few monolayer thicknesses as shown in Fig. \ref{Ek}. 
Since the $\Sigma_c$ valley is 6-fold degenerate, and the $K_c$ valley
is 3-fold degenerate, this results in a near 9-fold degeneracy
of the conduction band edge.
This increases the density of modes in the conduction band
by a factor of 9 from that of a single valley.
Furthermore, with increasing film thickness from 1L to 4L,
the splitting of the $\Sigma_c$
bands resulting from interlayer coupling is on the order of $k_BT$.
In MoS$_{2}$, the splitting at $\Sigma_{C}$ is 0.4 meV for the 2L and 40 meV
for the 4L structure.
The other materials show
similar magnitudes of the energy splitting as a function of thickness.
Therefore, 
the near-degeneracy of the bands at $\Sigma_c$ increases linearly with the film thickness, 
so that the number of modes per layer becomes relatively insensitive to the
layer thickness for few monolayer thicknesses.

The interlayer coupling of the out-of-plane d$_{z^{2}}$ and  p$_z$ orbitals 
result in the strongest interlayer hybridization and energy level 
splitting.\cite{TMDC_heterostructures_PRB13}
In MoS$_{2}$ the orbital composition of the conduction band $\Sigma_c$ valley 
is 
36$\%$ d$_{z^{2}}$,
22$\%$ d$_{xy}$, 
23$\%$ d$_{x^{2}-y^{2}}$, 
6\% p$_z$, and 
5\% p$_x$ and p$_y$. 
The d-orbital composition of the $K_c$ valley is 67\% d$_{z^2}$.
With increasing layer number, the $K_c$ valley splitting is larger than 
$k_BT$ so that the number of modes contributed by the $K_c$ valleys remains 3
independent of layer number. 
Thus, when the $\Sigma_c$ valley falls within $k_BT$ of the $K_c$
valley, its contribution to the density of modes dominates for few-layer
thicknesses.
Beyond 4 layers, the total splitting becomes
larger than $2k_BT$, and the number of accessible modes at $\Sigma_c$ no longer increases
linearly with thickness.

Beyond a monolayer, the valence band shifts to $\Gamma_v$ for MoS$_{2}$, MoSe$_{2}$ 
and WS$_{2}$.
The energy difference between $\Gamma_{v}$ and $K_{v}$ varies as a function of the film
thickness and material.
For MoS$_{2}$ the energy difference between $\Gamma_{v}$ and $K_{v}$ increases from
35 meV for the bilayer to 470 meV and 510 meV for the 3L and 4L structures respectively.
The near degeneracy of the $\Gamma_{v}$ and $K_{v}$ valleys 
leads to the largest p-type density of modes for 2L MoS$_{2}$.
For MoSe$_{2}$,
the $\Gamma_{v}$ and $K_{v}$ valleys are nearly degenerate above a single monolayer.
%
In WS$_{2}$, the energy difference of the $\Gamma_{v}$ and $K_{v}$ valleys 
decreases from 42 meV to 21 meV as the film thickness is increased from a bilayer
to four layers.
For WSe$_{2}$, the valence band maxima continues to reside at $K_{v}$ beyond a monolayer.
Once the valence band $K_v$ valleys begin to contribute in MoSe$_{2}$,  WS$_{2}$,
and WSe$_{2}$, the
density of modes per layer becomes relatively independent of layer thickness,
since there is little splitting of the $K_v$ valleys due to the interlayer
coupling.\cite{TMDC_heterostructures_PRB13}
The $K_v$ valley orbital composition contains no d$_{z^2}$
or $p_z$ components.
In MoS$_{2}$, the splitting varies from 0.2 meV for the 2L structure to
7.6 meV for the 4L structure.
The other materials show
similar magnitudes of the energy splitting as a function of thickness.
Thus, at room temperature,
the number of contributed modes per layer
within $k_BT$ of the $K_v$ valley minimum remains constant
for thicknesses in the range of one to four monolayers.

For the 8 layer WS$_2$ structure,
the conduction band $K_c$ and $\Sigma_c$ valleys are still
nearly degenerate. 
The $K_c$ valley lies 21 meV above the $\Sigma_c$ valley.
However, at both valleys,
the total splitting of the 8 bands contributed from the 8 layers is
much greater than $k_BT$ at room temperature.
At $\Sigma_c$, only two of the 8 bands are within 26 meV of the valley minimum.
The overall energy splitting of the 8 bands at $\Sigma_c$ is 193 meV.
In the valence band, the $K_v$ valley is 22 meV below the $\Gamma_v$ valley.
However, the 8 bands from the 8 monolayers are split over a total range of 180 meV,
and the second band is 40 meV below the $K_v$ valley maximum.
Thus, as the number of layers increase, the total energy splitting of the
bands contributed from each layer increases, and the number of 
modes per layer within $k_BT$ of the valley minimums decreases.

As the number of layers becomes macroscopic such that the crystal is periodic
in all three dimensions,
the total splitting of the bands evolves into the width of the dispersive band
along $k_z$ for the bulk crystal.
For bulk WS$_2$, the width of the band along the vertial $k_z$ direction
from  $\Sigma_c$ to $R$ at the top of the Brillouin zone is 208 meV which
is 15 meV larger than the total splitting of the 8 layer stack.
Furthermore, in the bulk, the $K_c$ valley is 126 meV above the $\Sigma_c$ minimum,
so the $K_c$ contributes no modes to the density of modes near the conduction
band edge.
In the valence band, the $K_v$ valley maximum is 225 meV below the $\Gamma_v$ 
maximum, so that the density of modes near the valence band edge, is entirely
from the $\Gamma_v$ valley.
The lack of valley near-degeneracy and the width of the bulk dispersive bands
along $k_z$, result in a minimum density of modes per layer near the band edges
compared to those of few layer structures.
The reduced number of modes per layer within $k_BT$ of the band edges
results in reduced per-layer values of the
thermoelectric figure of merit.

For both material systems Bi$_2$Te$_3$ and the semiconducting TMDCs, 
the enhancement of ZT results from the increased
degeneracy or near-degeneracy of the band edges. 
The origin and nature of the degeneracy is different.
In the Bi$_{2}$Te$_{3}$, the valence band edge becomes inverted into a ring
as a result of the coupling of the topological surface states.
In the TMDCs at few-layer thicknesses, different valleys become nearly degenerate.
In the conduction band,
the $\Sigma_c$ valleys become nearly degenerate 
with the $K_c$ valleys, and they contribute 6 more modes
to the 3 modes from the $K_c$ valleys. 
In the valence band,
the $K_v$ valleys become nearly degenerate with
the $\Gamma_v$, and they contribute 3 more modes.
Furthermore, because of the weak interlayer coupling at $K_v$ and $\Sigma_c$,
the additional bands from additional layers lie within $k_BT$ of the
band edges for few layers.
The increased band-edge degeneracy results in a sharper turn-on of the density of modes
and an increased value of ZT.

\section{SUMMARY}\label{sec:Summary}
The electronic structure of one to four monolayers of the semiconducting
transition metal dichalcogenides MoS$_{2}$, MoSe$_{2}$, WS$_{2}$ WSe$_{2}$
are calculated using DFT with spin-orbit coupling and the PBE functional.
Comparisons are made to results in the absence of spin-orbit coupling,
and the PBE results are compared to HSE calculations for MoS$_{2}$.
The peak n-type value of ZT increases by a factor of $6 - 8$ over the
bulk value for all materials.
Among the 4 materials and 4 thicknesses,
bilayer MoSe$_2$ gives the maximum n-type ZT value of 2.4.
The peak p-type value of ZT increases by a factor of $5 - 14$ over the
bulk value for all materials.
The maximum p-type ZT value of 1.2 occurs for bilayer MoS$_2$.
%
%
The maximum power factor generally occurs for a different 
layer thickness and at a more degenerate Fermi level
than the maximum value of ZT.
This difference can be explained by the increased
electrical thermal conductivity at the Fermi level
corresponding to the maximum power factor.
For all materials, the maximum value of ZT coincides with the
sharpest turn-on of the density of modes distribution at the band 
edge.
The sharper turn-on is driven by 
the near valley degeneracy 
of the conduction band $K_c$ and $\Sigma_c$ valleys
and the valence band $\Gamma_v$ and $K_v$ valleys.
For few layer structures, the degeneracy is enhanced by the 
weak interlayer coupling at the $\Sigma_c$ and $K_v$ valleys.

\noindent
\begin{acknowledgements}
This work is supported in part by the National Science Foundation (NSF)
Grant Nos. 1124733 and 1128304
and the Semiconductor Research Corporation (SRC) Nanoelectronic Research Initiative
as a part of the Nanoelectronics for 2020
and Beyond (NEB-2020) program, FAME, one of six centers of STARnet, a 
Semiconductor Research Corporation program sponsored by MARCO and DARPA, and
the University Grant Council (Contract No. AoE/P-04/08) of the Government of HKSA (FZ).
This work used the Extreme Science and Engineering Discovery Environment (XSEDE), 
which is supported by National Science Foundation grant number OCI-1053575.

\end{acknowledgements}

\begin{center}

\mbox{ }
\newpage
\begin{figurehere}
\includegraphics[width=5in]{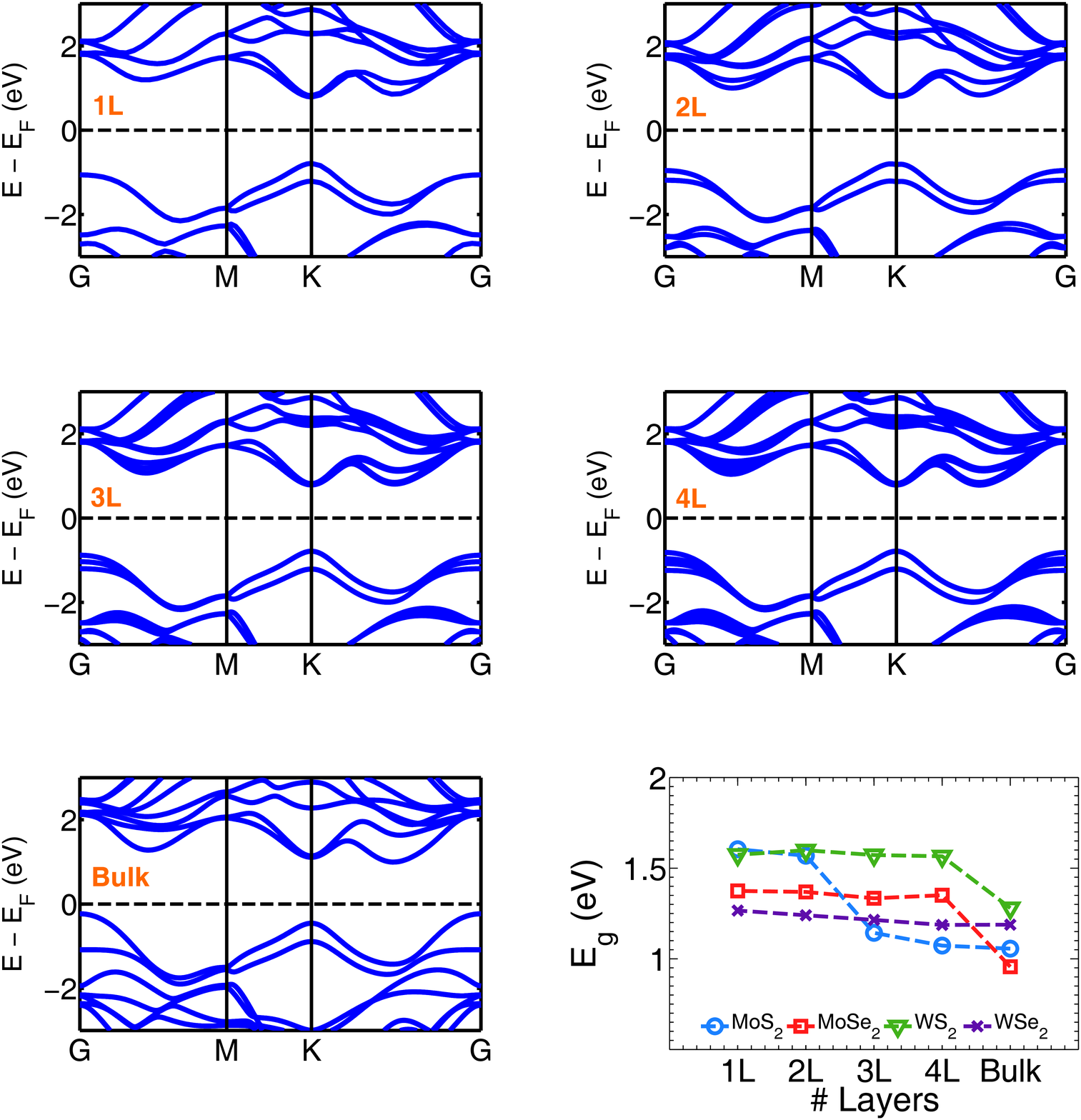}
\caption{ \emph{Ab-initio} calculated band structures of WS$_{2}$: 1L, 2L, 3L, 
4L and bulk. 
The bottom right panel illustrates the variation of the band gap of the TMDC materials 
as a function of the number of layers.
\label{Ek}
}
\end{figurehere}

\mbox{ }
\newpage
\begin{figurehere}
\includegraphics[width=5in]{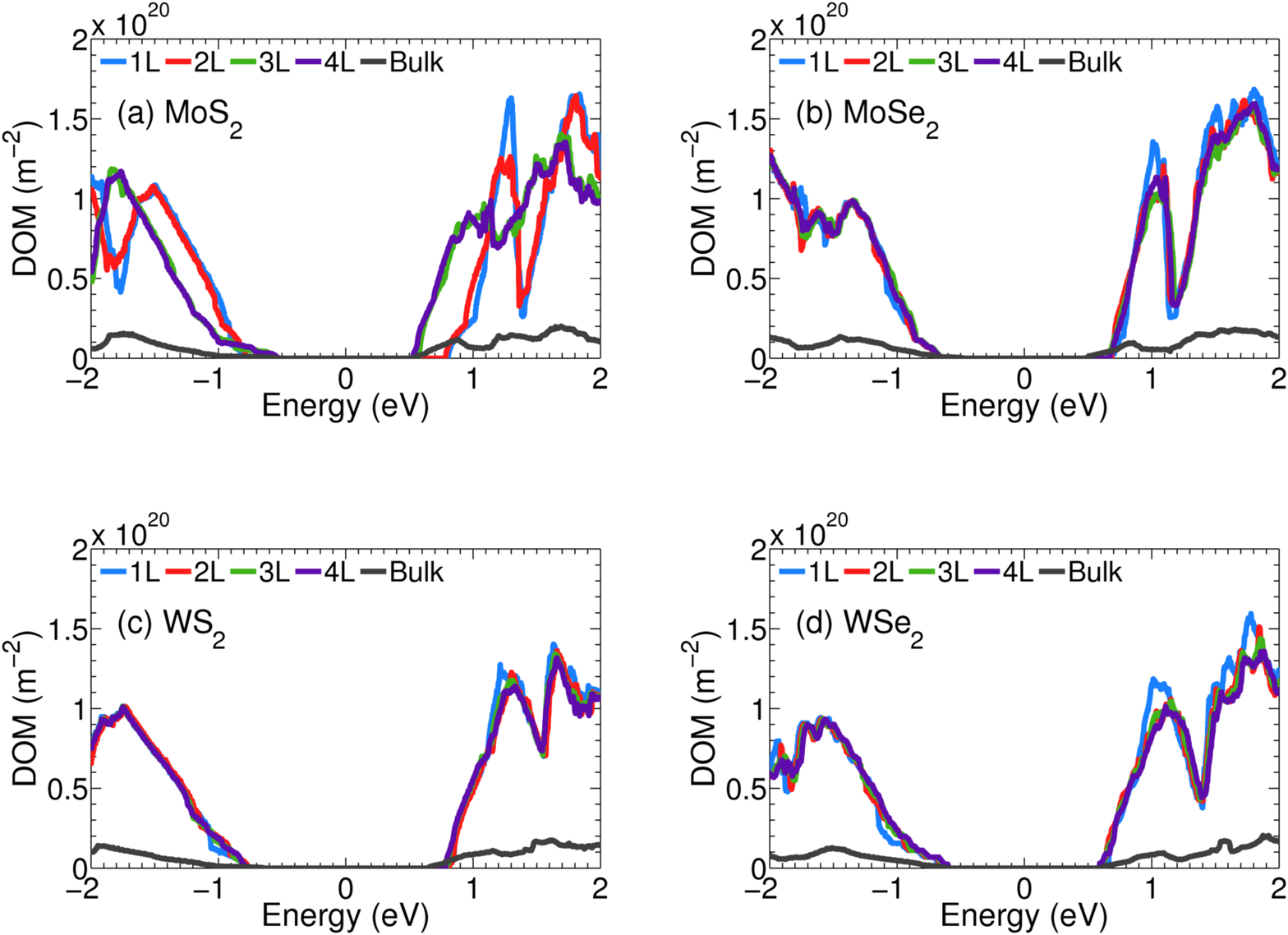}
\caption{  (Color online) Distribution of modes per 
unit area versus energy for (a) MoS$_{2}$, (b) MoSe$_{2}$,
(c) WS$_{2}$  and (d) WSe$_{2}$ for bulk (black),
1L (blue), 2L (red), 3L (green) and 4L (purple) structures. 
The midgap energy is set to E=0.  }
\label{DOM}
\end{figurehere}

\mbox{ }
\newpage
\begin{figurehere}
\includegraphics[width=5in]{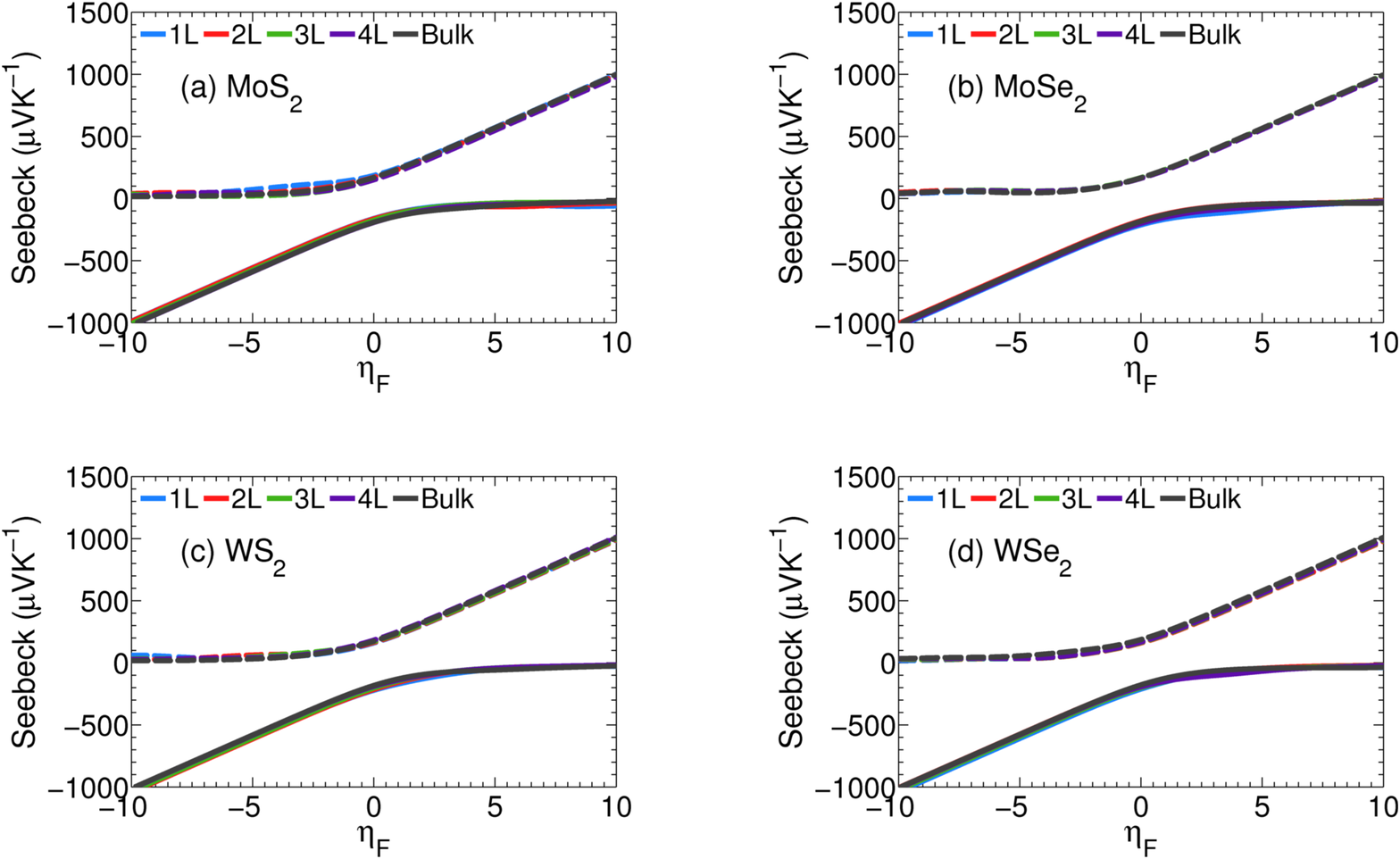}
\caption{  (Color online) Seebeck coefficient at 300K for (a) MoS$_{2}$,
(b) MoSe$_{2}$, (c) WS$_{2}$ and (d) WSe$_{2}$ for bulk (black),
1L (blue), 2L (red), 3L (green) and 4L (purple) 
 structures.  The n-type Seebeck coefficients are plotted with a solid
 line and p-type coefficients with a broken line 
 as a function of the reduced Fermi energy, $\eta_{F}$.  }
\label{Seebeck}
\end{figurehere}

\mbox{ }
\newpage
\begin{figurehere}
\includegraphics[width=5in]{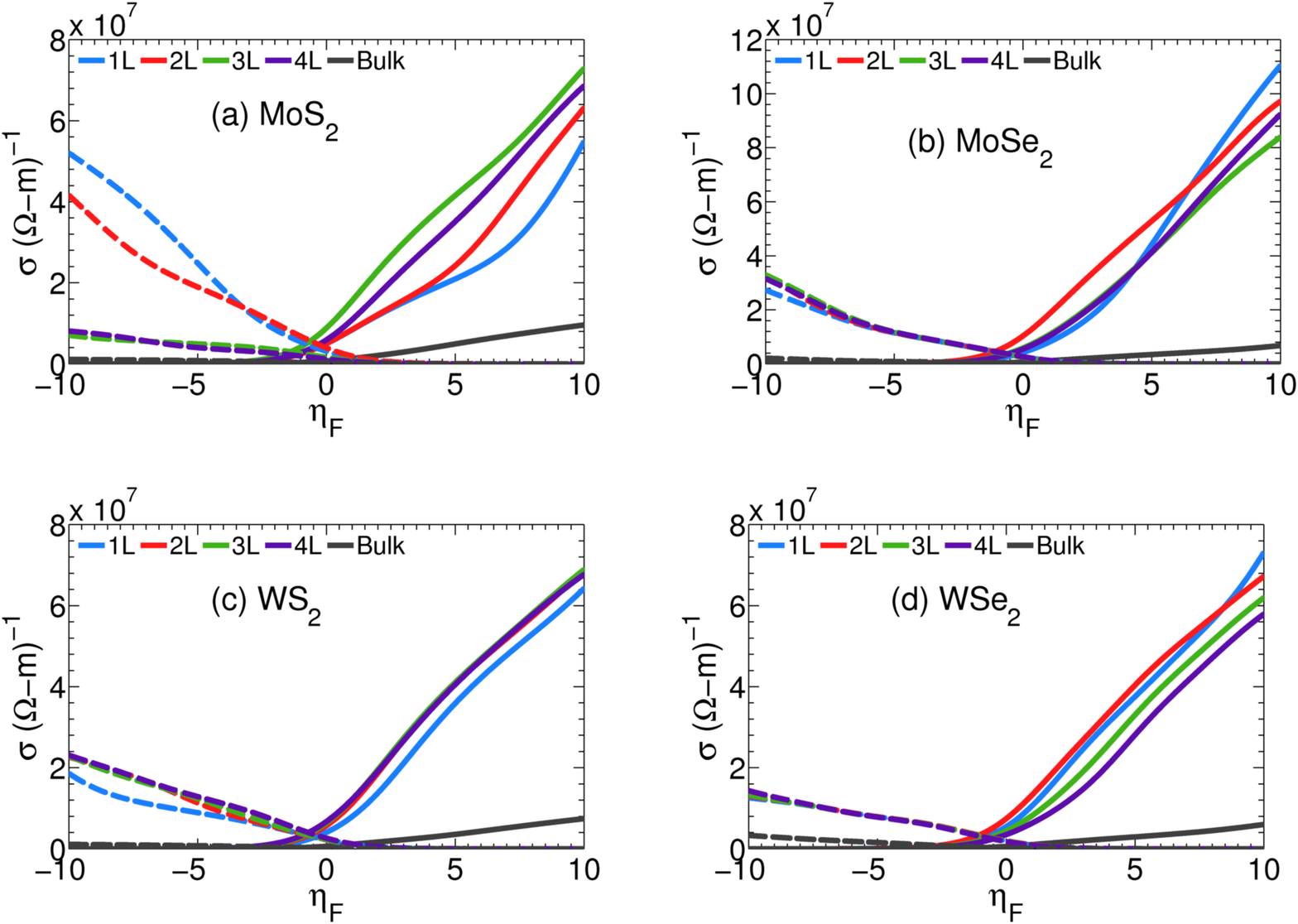}
\caption{ (Color online) Electrical conductivity, $\sigma$,
 at 300K for (a) MoS$_{2}$,
(b) MoSe$_{2}$, (c) WS$_{2}$ and (d) WSe$_{2}$ for 1L (blue), 
2L (red), 3L (green) and 4L (purple) and bulk (black) 
 structures.  The n-type electrical conductivity is plotted with a solid
 line and p-type conductivity with a broken line 
 as a function of the reduced Fermi energy, $\eta_{F}$.  }
\label{El_Conductivity}
\end{figurehere}

\mbox{ }
\newpage
\begin{figurehere}
\includegraphics[width=5in]{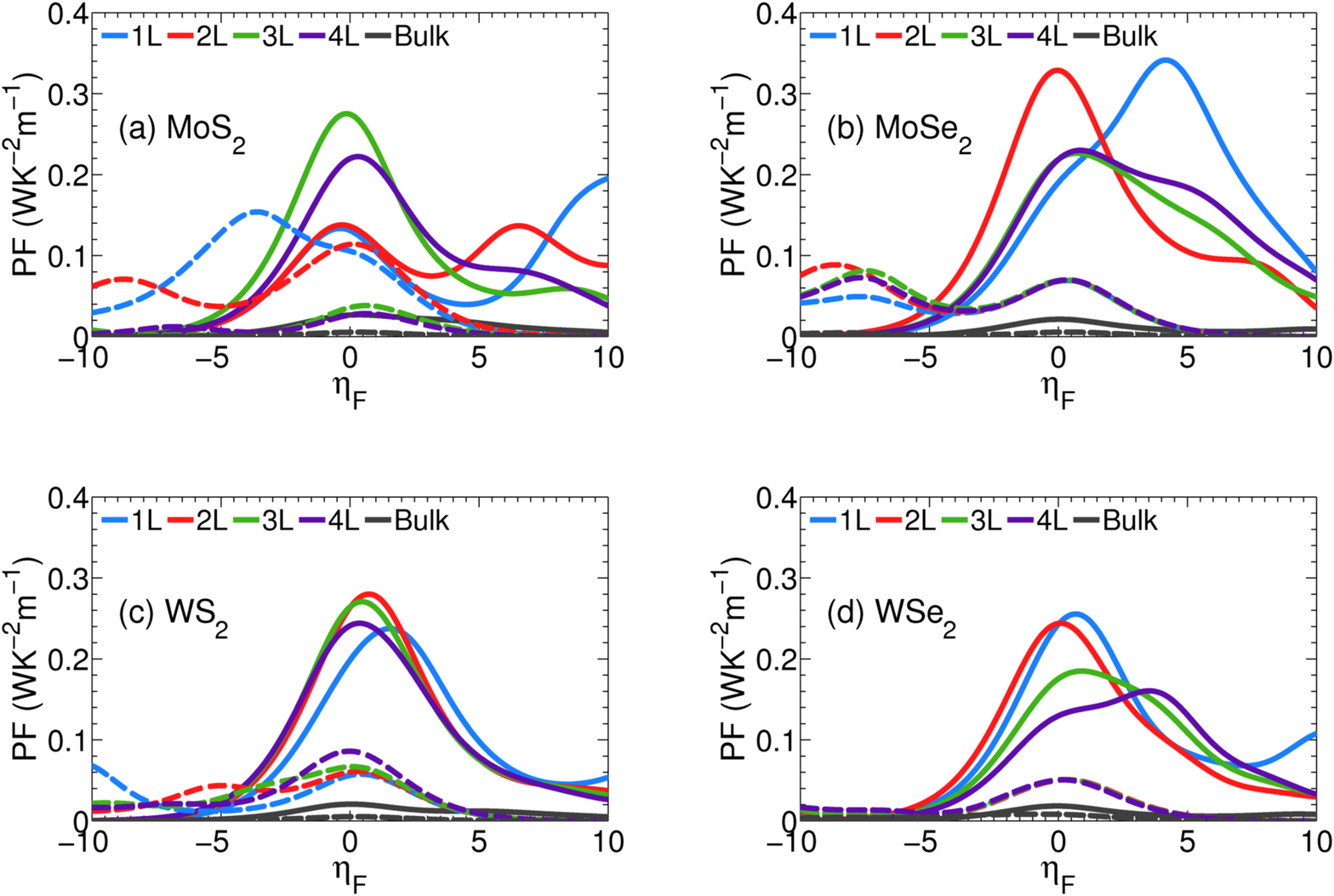}
\caption{  (Color online) Power factor (PF) at 300K for (a) MoS$_{2}$,
(b) MoSe$_{2}$, (c) WS$_{2}$ and (d) WSe$_{2}$ for bulk (black),
1L (blue), 2L (red), 3L (green) and 4L (purple) 
 structures. The n-type power factors are plotted with a solid
 line and p-type PFs with a broken line
 as a function of the reduced Fermi energy, $\eta_{F}$.  }
\label{PF}
\end{figurehere}

\mbox{ }
\newpage
\begin{figurehere}
\includegraphics[width=5in]{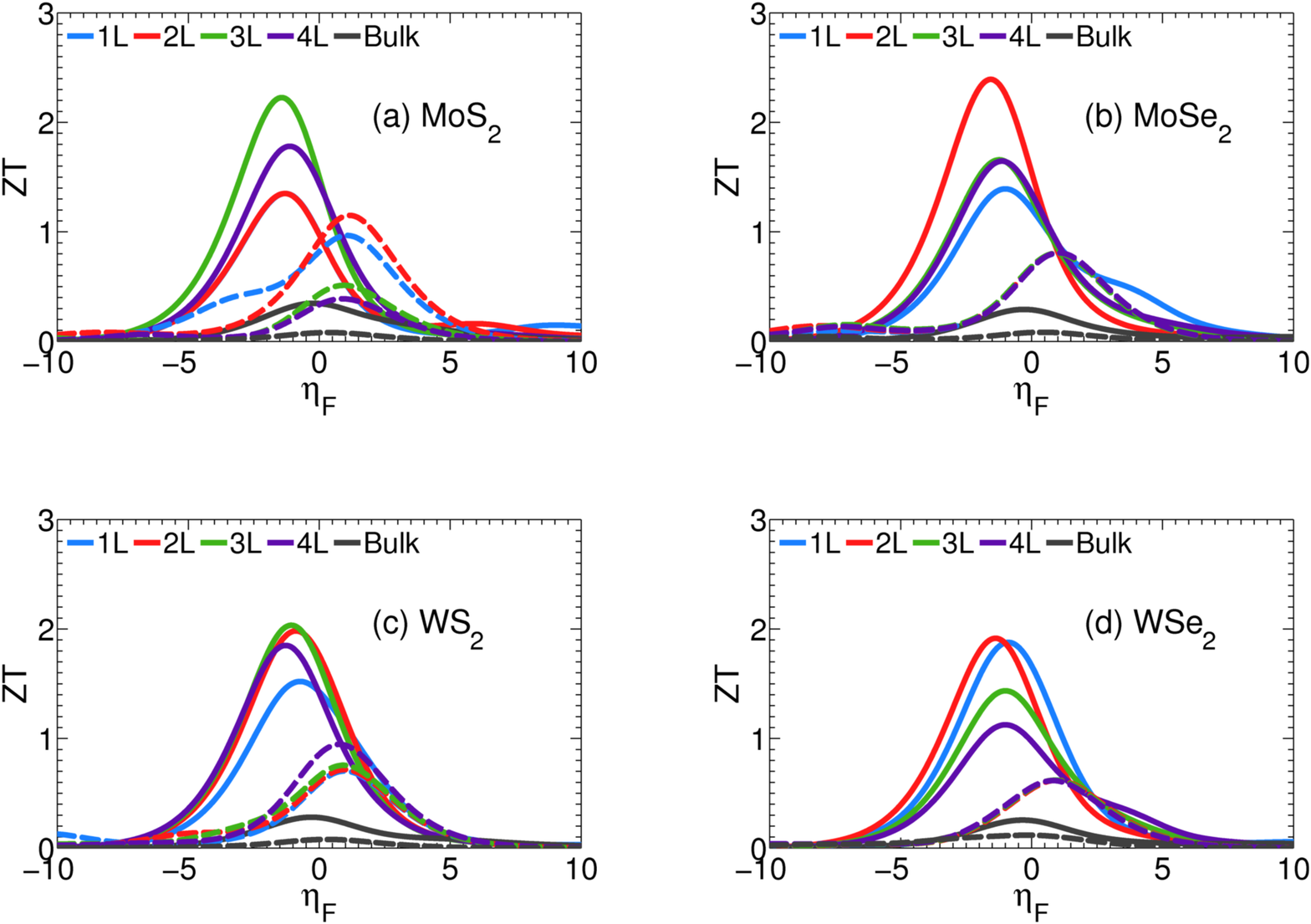}
 \caption{ (Color online) ZT at 300K for (a) MoS$_{2}$, (b) MoSe$_{2}$,
(c) WS$_{2}$ and (d) WSe$_{2}$ for  bulk (black),
1L (blue), 2L (red), 3L (green) and 4L (purple)  
structures.  The n-type ZT is plotted with a solid
 line and p-type ZT with a broken line
 as a function of the reduced Fermi energy, $\eta_{F}$.  }
 \label{ZT}
\end{figurehere}

\mbox{ }
\newpage
\begin{figurehere}
\includegraphics[width=5in]{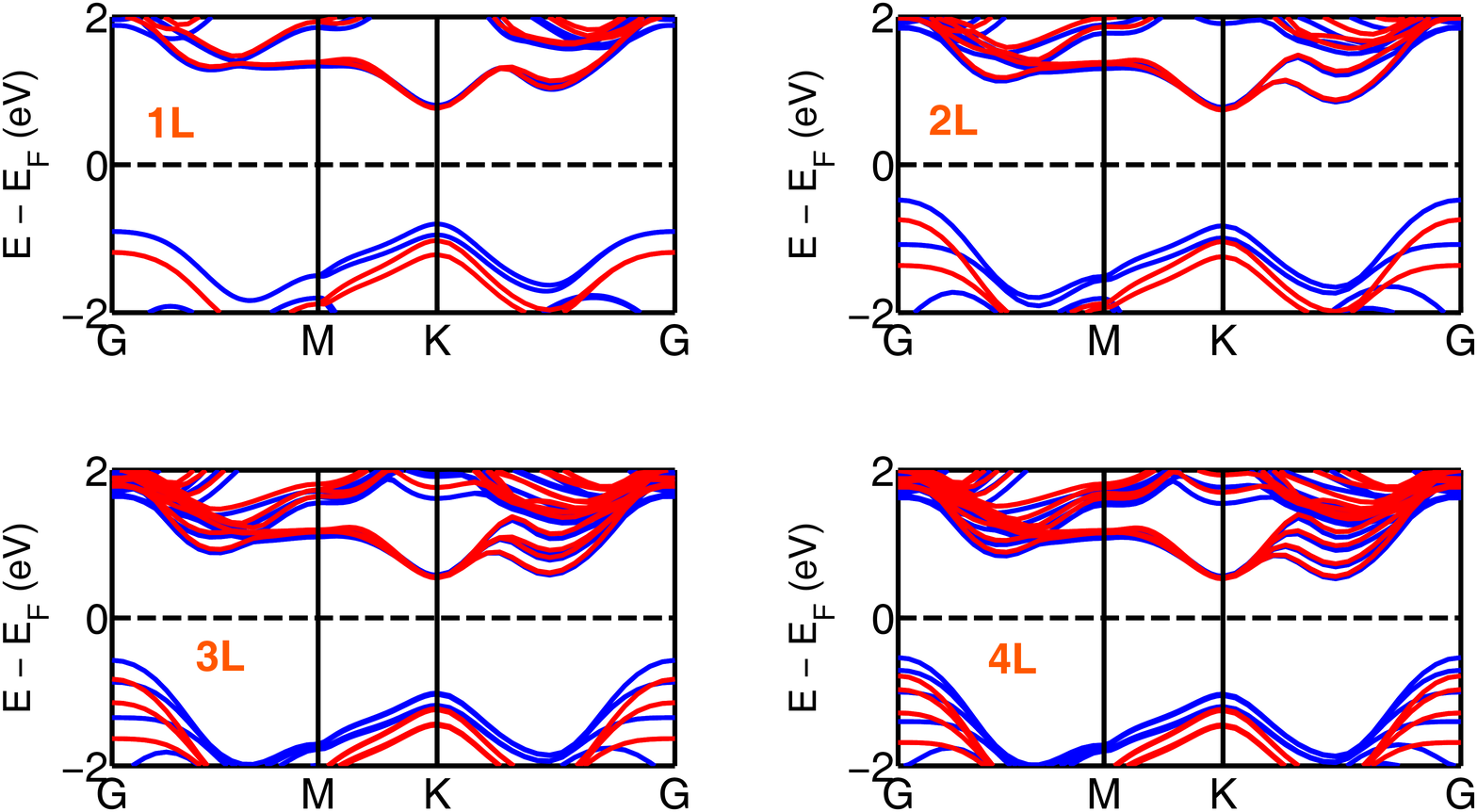}
\caption{ \emph{Ab-initio} calculated electronic structure of MoS$_{2}$: 
1L, 2L, 3L and 4L structures using a PBE (blue) and hybrid HSE (red) functional. 
The HSE functional provides a correction to the underestimated PBE 
bandgap while the salient features of the electronic structure that 
would affect the density-of-modes calculation remain the same.
\label{MoS2_PBE_HSE}
}

\end{figurehere}

\mbox{ }
\newpage
\begin{figurehere}
\includegraphics[width=5in]{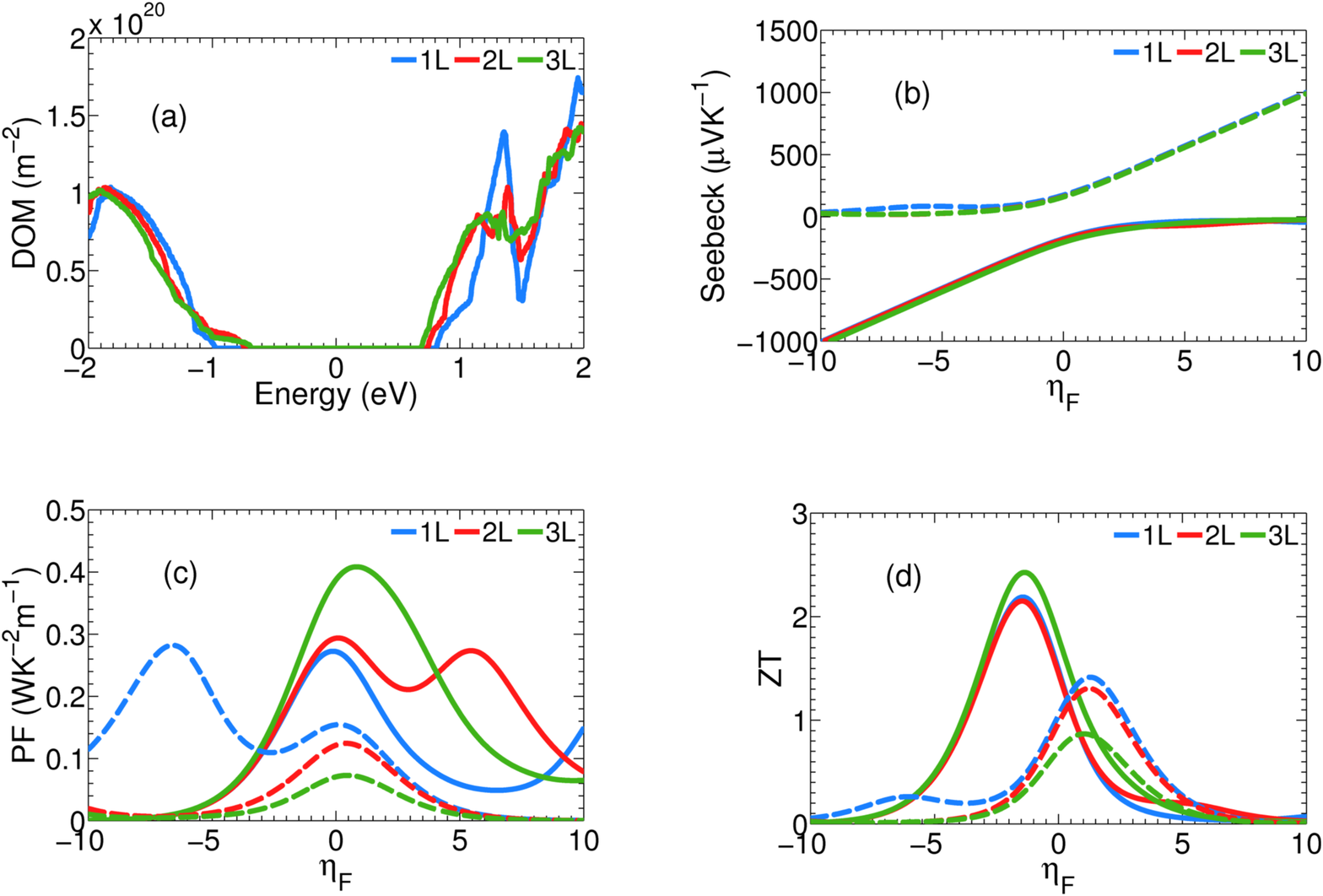}
\caption{(Color online) HSE calculation of the (a) density-of-modes,
(b) Seebeck coefficient, (c) Power factor and (d) ZT 
1L (blue), 2L (red), 3L (green) MoS$_{2}$. The n-type thermoelectric parameters
are plotted with a solid line and the p-type parameters are plotted
with a broken line as a function of the reduced Fermi energy, $\eta_{F}$.}
\label{MoS2_HSE}

\end{figurehere}

\mbox{ }
\newpage
\begin{figurehere}
\includegraphics[width=5in]{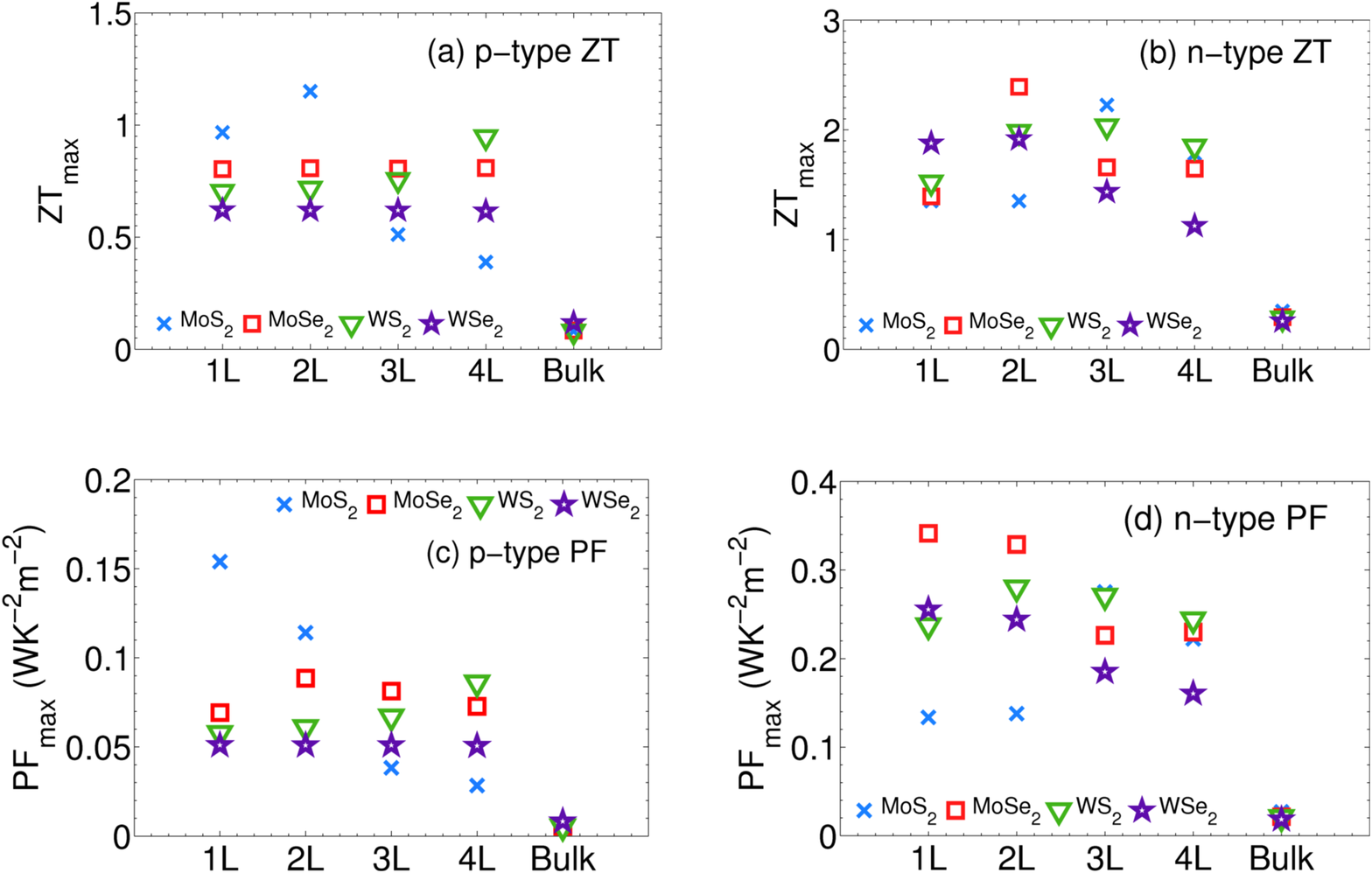}
 \caption{(Color online) Maximum thermoelectric performance for 1L (blue), 2L (red), 3L (green),
 4L (purple) and bulk (black) MoS$_{2}$, MoSe$_{2}$, WS$_{2}$, WSe$_{2}$ at 
 300K: (a) Maximum p-type ZT, 
(b) Maximum n-type ZT, (c) Maximum p-type power factor, (d) Maximum n-type power factor.
}
\label{ZT_PF_dimension}

\end{figurehere}

\mbox{ }
\newpage
\begin{figurehere}
\includegraphics[width=5in]{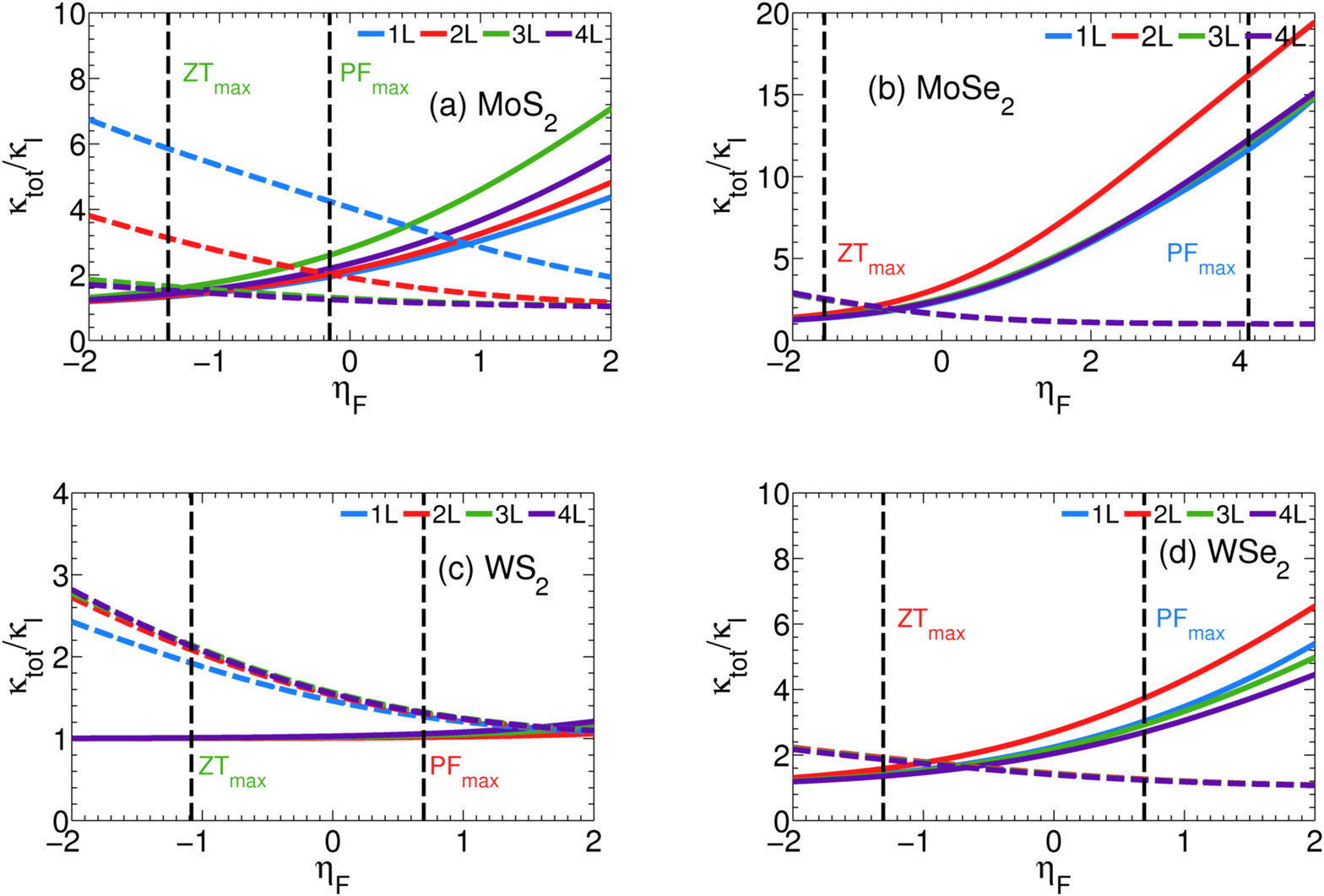}
\caption{ (Color online) 
Ratio of total thermal conductivity ($\kappa_{l}$+$\kappa_{e}$) over
the lattice thermal conductivity ($\kappa_{l}$) at 300K for 
(a) MoS$_{2}$, (b) MoSe$_{2}$, (c) WS$_{2}$, 
(d) WSe$_{2}$ for 1L (blue), 2L (red), 3L (green) and 4L (purple)  
structures.  
The n-type ratio is plotted with a solid
line and p-type ratio with a broken line
as a function of the reduced Fermi energy, $\eta_{F}$.
The two vertical dashed lines show the reduced Fermi level position
at which the maximum n-type power factor and ZT occur.
}
\label{kappa_eff}
\end{figurehere}

\newpage
\begin{figurehere}
\includegraphics[width=5in]{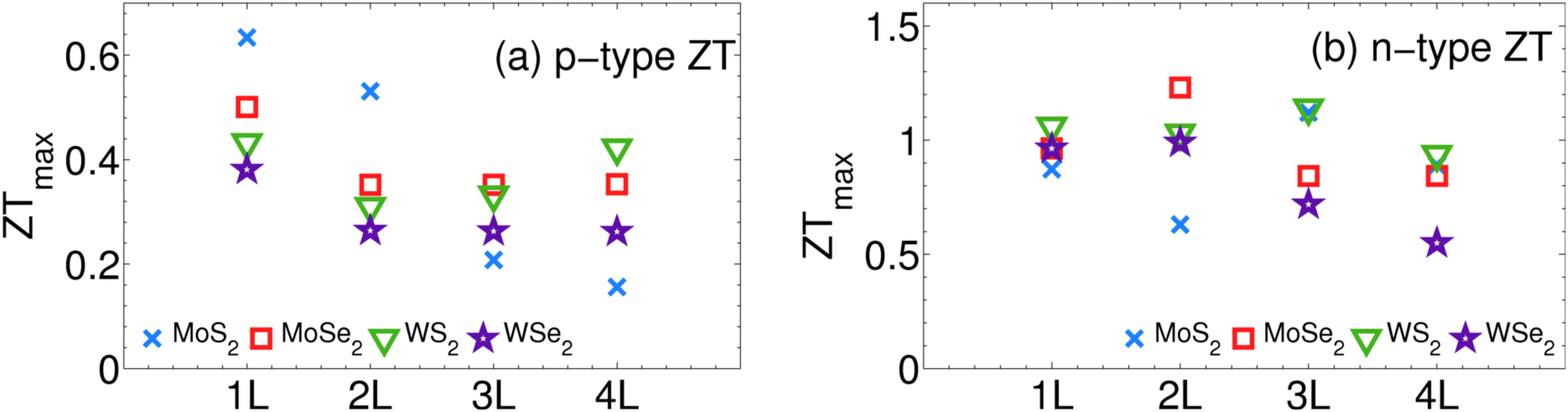}
 \caption{(Color online) Maximum ZT for (a) p-type and (b) n-type of 
MoS$_{2}$, MoSe$_{2}$, WS$_{2}$, WSe$_{2}$ at 
 300K for 1L (blue), 2L (red), 3L (green),
 4L (purple) structures accounting for thickness-dependent 
lattice thermal conductivity.  
$\kappa_{l}$ =34.5 Wm$^{-1}K^{-1}$ used for the 1L structures and 
 $\kappa_{l}$=52 Wm$^{-1}K^{-1}$ used for the few-layer structures.}
\label{ZT_kappa_thickness}
\end{figurehere}

\mbox{ }
\newpage
\begin{figurehere}
\includegraphics[width=5in]{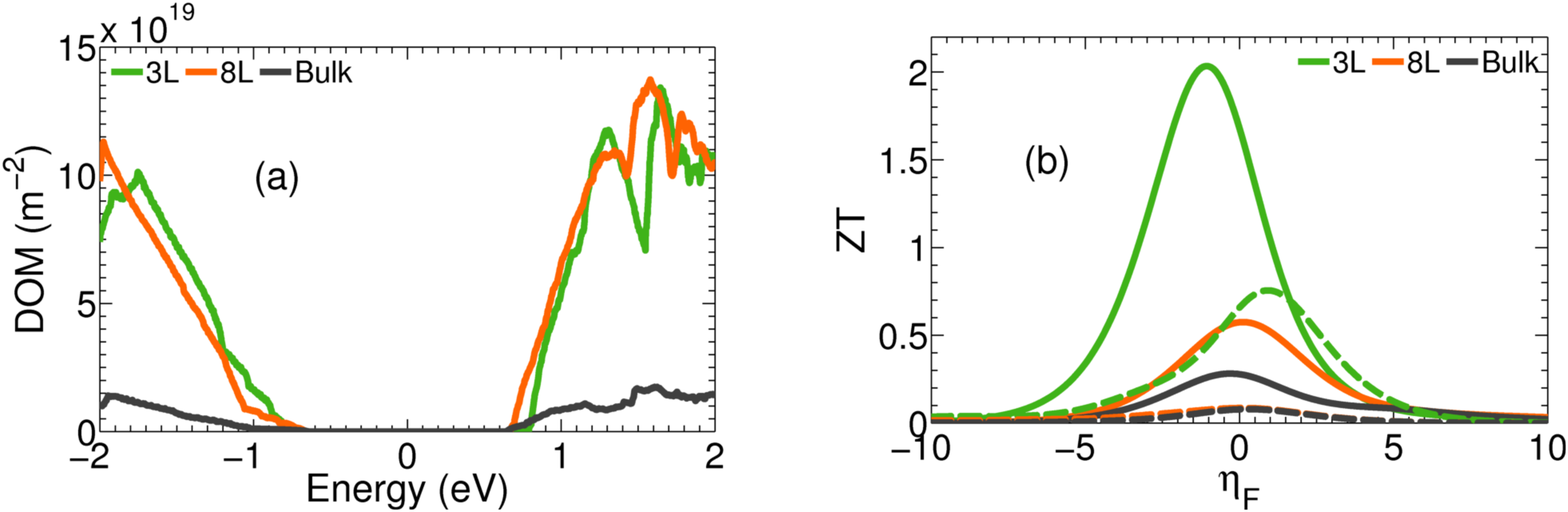}
 \caption{(Color online) (a) Density of modes and (b) ZT as a function
of the reduced Fermi level for 3L (green), 8L (orange) and bulk (bulk) WS$_{2}$.  
}
\label{WS2_DOM_ZT_bulklike}

\end{figurehere}

\mbox{ }
\newpage
\begin{figurehere}
\includegraphics[width=5in]{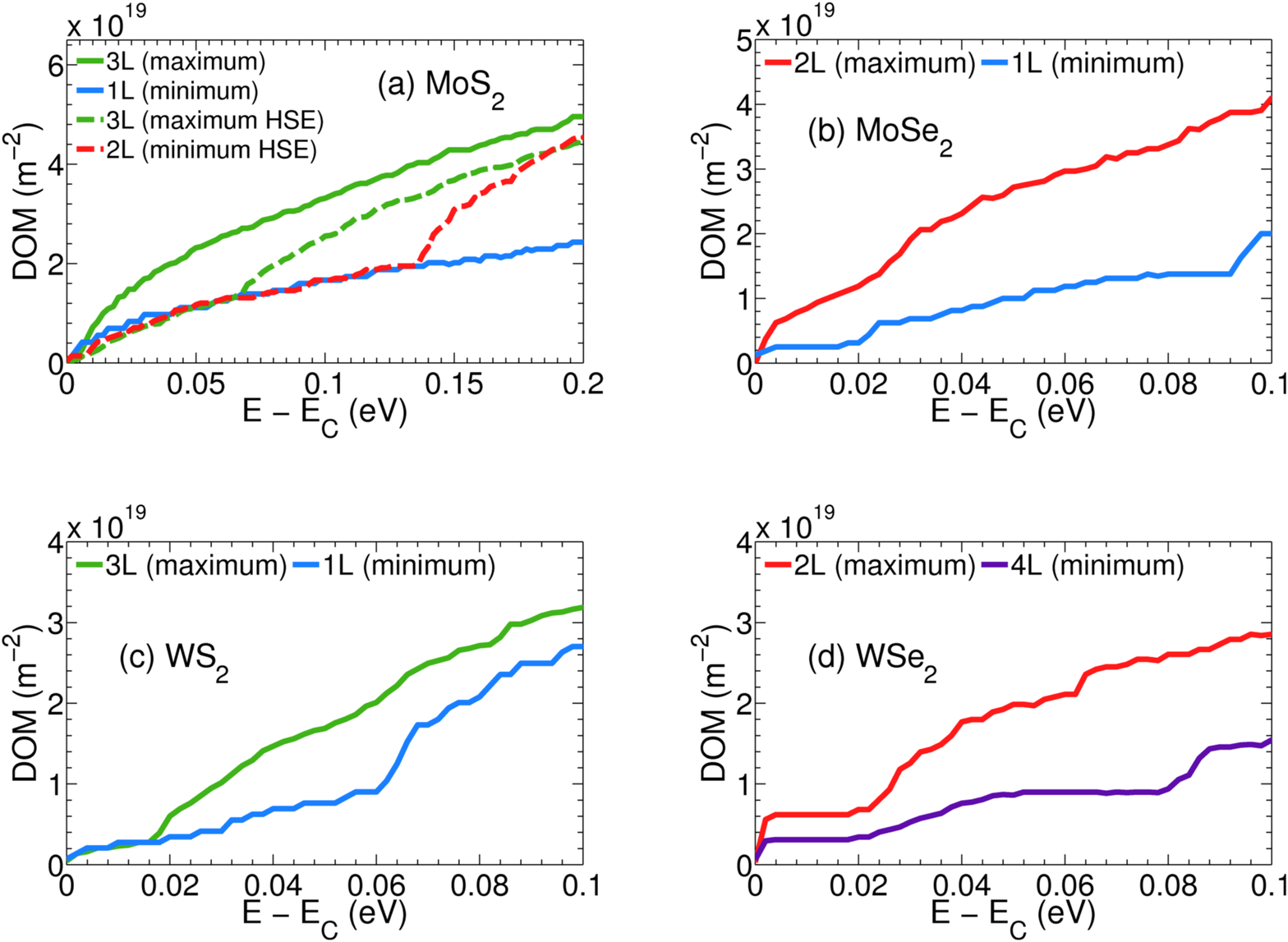}
 \caption{(Color online) Conduction band density of modes (DOM)
 for (a) MoS$_{2}$, (b) MoSe$_{2}$, (c) WS$_{2}$ and (d) WSe$_{2}$
at film thicknesses where the maximum and the minimum ZT occurs with
respect to the energy away from the conduction band edge, E$_{C}$.  
}
\label{DOM_CB}

\end{figurehere}

\mbox{ }
\newpage
%
\end{center}

\newpage
\begin{center}
\vspace{0.5in}
\begin{table*}
\begin{tabular}{p{1cm} p{1.5cm} p{1.5cm} p{1.5cm} p{1.5cm} p{1.5cm} p{1.5cm} p{1.5cm} p{1.5cm}}
\hline\hline
& $a_{0}$($\AA$) &  $c_{0}$($\AA$)  & $z$ &  $a_{0}^{expt}(\AA$) & $c_{0}^{expt}(\AA$) & $z^{expt}$ & $E_{g}$(eV) & $E_{g}^{expt}$(eV) \\[0.5ex]  \hline
MoS$_{2}$ & 3.179 & 12.729 &  0.627 & 3.160  & 12.290  & 0.629 & 1.060 & 1.29 \\ [0.5ex]
MoSe$_{2}$& 3.309 & 13.323 &  0.624 & 3.289  & 12.927  & 0.621 & 0.959 & 1.09 \\ [0.5ex]
WS$_{2}$  & 3.183 & 13.131 &  0.630 & 3.150  & 12.320  & 0.622 & 1.283 & 1.35 \\ [0.5ex]
WSe$_{2}$ & 3.319 & 13.729 &  0.627 & 3.282  & 12.960  & 0.621 & 1.188 & 1.20 \\ [0.5ex]
\hline 
\end{tabular}
\caption{Calculated properties of bulk TMDC materials: 
lattice constant $a_{0}$, c-axis lattice constant $c_{0}$, z-parameter $z$,
and bandgap $E_{g}$(eV).  Experimental values
\cite{MoS2_MoSe2_structure_Jellinek, WS2_WSe2_structure_Jellinek, Kam_bulk_Egap_JPhysChem} 
have been included for comparison.}
\label{tab:params} 
\end{table*}
\end{center}

\begin{flushleft}
\vspace{0.2in}
\begin{table*}
\begin{tabular} {c | c |  p{1.7cm} p{1.7cm} p{1.7cm} p{1.7cm} | p{1.7cm} p{1.7cm} p{1.7cm} p{1.7cm}}
\hline\hline
Structure & Point &  MoS$_{2}$  & MoSe$_{2}$ & WS$_{2}$ & WSe$_{2}$ &  MoS$_{2}$  & MoSe$_{2}$ & WS$_{2}$ & WSe$_{2}$   \\[0.5ex]  \hline
& &\multicolumn{4}{c|}{\bf{Hole Effective Mass (m$_{0}$)}} & \multicolumn{4}{c}{\bf{Electron Effective Mass (m$_{0}$)}} \\[0.5ex] \hline
1L & K$_{l}$ & 0.543 & 0.578 & 0.339 & 0.341 & 0.506 & 0.502 & 0.349 &   0.345\\ [0.5ex]
    & K$_{t}$ & 0.546 & 0.588 &0.339 & 0.348 & 0.504 & 0.503 & 0.347 &   0.345\\ [0.5ex] \hline

2L & $\Gamma$ & 1.039 & 1.430 & 1.239 & 1.322  & - & - & - & -  \\ [0.5ex]
    & K$_{l}$ & 0.548 & 0.595 & 0.345 & 0.349 & 0.521 & 0.539 & 0.359 &  0.411   \\ [0.5ex]
    & K$_{t}$ &  0.546 & 0.596 & 0.346 & 0.348 & 0.510 & 0.539 & 0.359 &  0.412 \\ [0.5ex] \hline

3L & $\Gamma$ & 1.239 & 1.432 & 1.246 & 1.382 & - & - & - & - \\ [0.5ex]
    & K$_{t}$ & 0.549 & 0.602 & 0.366 & 0.368 & 0.559 & 0.544 & 0.376 & 0.434 \\ [0.5ex]
    & K$_{t}$ & 0.548 & 0.604 & 0.366 & 0.368 & 0.559 & 0.544 & 0.377 & 0.434\\ [0.5ex] \hline

4L & $\Gamma$ & 1.239 & 1.433 & 1.351 & 1.432 & - & - & - & - \\ [0.5ex]
    & K$_{l}$  & 0.548 & 0.604 & 0.366 & 0.367 & 0.554 & 0.542 & 0.376 & 0.435 \\ [0.5ex]
    & K$_{t}$  & 0.546 & 0.604 & 0.366 & 0.368  & 0.559 & 0.549 & 0.377 & 0.434 \\ [0.5ex] \hline

Bulk & $\Gamma$ & 0.838 & 0.973 & 0.832 & 0.997 & - & - & - & -\\ [0.5ex]
   & $\Sigma_{l}$  & - & - & - & - & 0.590 & 0.521 & 0.569 & 0.489    \\ [0.5ex]
   & $\Sigma_{t}$  & - & - & - & - &  0.845 & 0.776 & 0.665 &    0.643  \\ [0.5ex]
\hline
\end{tabular}
\caption{\emph{Ab-initio} calculations of the hole and electron effective masses
at the valence band maxima and conduction band minima respectively for each structure in units of
the free electron mass (m$_{0}$). The subscripts $l$ and $t$ refer to the masses calculated at 
the symmetry point along the longitudinal and the transverse directions.}
\label{tab:Eff_Mass} 
\end{table*}
\end{flushleft}

\begin{flushleft}
\vspace{0.2in}
\begin{table*}
\begin{tabular} {c | c |  p{1.5cm} p{1.5cm} p{1.5cm} p{1.5cm}| p{1.5cm} p{1.5cm} p{1.5cm} p{1.5cm}}
\hline\hline
Structure & Transition &  MoS$_{2}$  & MoSe$_{2}$ & WS$_{2}$ & WSe$_{2}$ &  MoS$_{2}$  & MoSe$_{2}$ & WS$_{2}$ & WSe$_{2}$\\[0.5ex]  \hline
& &\multicolumn{4}{c|}{\bf{Calculated (eV)}} & \multicolumn{4}{c}{\bf{Experimental (eV)}} \\[0.5ex] \hline
1L &$\Gamma_{v}$ to K$_{c}$ & 1.705 & 1.768 & 1.849 & 1.776 & - & - & - & - \\ [0.5ex] 
   &$\Gamma_{v}$ to $\Sigma_{c}$ & 1.922 & 1.862 & 1.929 & 1.806  & - & - & - & - \\ [0.5ex] 
   & K$_{v1}$ to K$_{c}$ & \bf{1.600} & \bf{1.375} & \bf{1.573} & \bf{1.254} & 1.900 & 1.660 & 1.950 & 1.640 \\ [0.5ex]
   &K$_{v2}$ to K$_{c}$ & 1.750 & 1.556 & 1.973 & 1.715 & 2.050 & 1.850 & 2.360 & 2.040 \\ [0.5ex] \hline

2L & $\Gamma_{v}$ to K$_{c}$ & \bf{1.564} & \bf{1.368} & \bf{1.507} & 1.586 & 1.600 & - & 1.730 & - \\ [0.5ex]
   &$\Gamma_{v}$ to $\Sigma_{c}$ & 1.775 & 1.373 & 1.542  & 1.562  & - & - & - & - \\ [0.5ex] 
    & K$_{v1}$ to K$_{c}$ & 1.600 & 1.373 & 1.549 & 1.269 & 1.880 & - & 1.910 & 1.590 \\ [0.5ex]
    & K$_{v2}$ to K$_{c}$ &  1.760 & 1.556 & 1.977 & 1.788 & 2.050 & - & 2.340 & 2.000 \\ [0.5ex] \hline

3L & $\Gamma_{v}$ to K$_{c}$ & \bf{1.150} & \bf{1.334} & \bf{1.458} & 1.586 & - & - & - & -  \\ [0.5ex]
   &$\Gamma_{v}$ to $\Sigma_{c}$ & 1.171 & 1.372 & 1.482 & 1.508 & - & - & - & - \\ [0.5ex] 
   & K$_{v1}$ to K$_{c}$ & 1.620 & 1.376 & 1.485 & 1.265 & - & - & - & - \\ [0.5ex]
    & K$_{v2}$ to K$_{c}$ & 1.780 & 1.564 & 1.873 & 1.783 & - & - & - & - \\ [0.5ex] \hline

4L & $\Gamma_{v}$ to K$_{c}$ & \bf{1.120} & \bf{1.351} & \bf{1.438} & 1.546 & - & - & - & - \\ [0.5ex]
    &$\Gamma_{v}$ to $\Sigma_{c}$ & 1.139 & 1.374  & 1.439 & 1.434  & - & - & - & - \\ [0.5ex] 
    & K$_{v1}$ to K$_{c}$  & 1.630 & 1.356 & 1.459 & 1.259 & - & - & - & -  \\ [0.5ex]
    & K$_{v2}$ to K$_{c}$  & 1.780 & 1.574 & 1.877 & 1.753  & - & - & - & -  \\ [0.5ex] \hline

Bulk & $\Gamma_{v}$ to $\Sigma_{c}$ & \bf{1.060} & \bf{0.959} & \bf{1.283} & \bf{1.188} & 1.290 & 1.090 & 1.350 & 1.200 \\ [0.5ex]
   & K$_{v1}$ to K$_{c}$  & 1.590 & 1.349 & 1.453 & 1.258 &  1.880 & 1.350 &  1.880 & 1.580 \\ [0.5ex]
   & K$_{v2}$ to K$_{c}$  & 1.780 & 1.588 & 1.889 & 1.737 &  2.060 & 1.380 & 2.320 & 1.950 \\ [0.5ex]
\hline
\end{tabular}
\caption{ \emph{Ab-initio} calculations of the bandgap energies and energy transitions
between the valence ($v$) and conduction ($c$) band valleys for each structure and material.  The splitting of
the valence band at the $K$-point due to spin-orbit coupling 
and the inter-layer interactions are
denoted as $K_{v1}$ and $K_{v2}$.  
$\Sigma$ is the mid-point between $\Gamma$ and $K$.  The bandgap
at each dimension is highlighted in bold text.  
 Experimental values when available
\cite{MoS2_MoSe2_structure_Jellinek, WS2_WSe2_structure_Jellinek, Kam_bulk_Egap_JPhysChem, Cui_WS2_NatureSciRep_13}
 have been included for comparison.
\\
NOTE: The band gap of 2L, 3L and 4L WSe$_{2}$ occurs between the K$_{v1}$ and $\Sigma_{C}$.  The 2L, 3L and 4L 
band gaps of WSe$_{2}$ are 1.216 eV, 1.1594 eV and 1.1345 eV respectively.}
\label{tab:Energy_gaps}
\end{table*}
\end{flushleft}

\begin{flushleft}
\vspace{0.2in}
\begin{table*}
\begin{tabular}{c c p{2.2cm} p{2.2cm} p{2.2cm} p{2.2cm} c}
\hline\hline
& Temperature & \multicolumn{1}{c}{1L} &  \multicolumn{1}{c}{2L}  & \multicolumn{1}{c}{3L} &  \multicolumn{1}{c}{4L} & \multicolumn{1}{c}{Bulk}  \\[0.5ex]  \hline
& & \multicolumn{5}{c}{\bf{Maximum n-type (p-type) Power Factor ($WK^{-2}m^{-2}$)}} \\[0.5ex] \hline
MoS$_{2}$ & 300K & .130 (.150) & .140 (.110) &  \textbf{.280} (.041) & .220 (.031) & .0320 (.010) \\ [0.5ex]
	  & 150K & .093 (.072) & .093 (.071) &  \textbf{.190} (.032) & .120 (.024) & .012 (.0042) \\ [0.5ex]
	  & 77K & .072 (.043) & .072 (.053) &  \textbf{.13} (.021) & .063 (.022) & .012 (.0031)   \\[0.5ex] \hline
MoSe$_{2}$ & 300K & \textbf{.340} (.071) & .330 (.094) &  .230 (.082) & .230 (.071) & .022 (.0061) \\ [0.5ex]
          & 150K & \textbf{.151} (.050) & .200 (.051) &  .100 (.051) & .100 (.052) & .013 (.004) \\ [0.5ex]
	  & 77K & \textbf{.062} (.031) & .120 (.032) &  .062 (.031) & .052 (.032) & .013 (.0032) \\ [0.5ex] \hline
WS$_{2}$  & 300K & .240 (.062) & \textbf{.280} (.061) &  .270 (.071) & .240 (.092)  & .022 (.0052) \\ [0.5ex]
          & 150K & .110 (.042) & \textbf{.160} (.042) &  .150 (.041) & .130 (.051) & .010 (.0043) \\ [0.5ex]
	  & 77K & .051 (.031) & \textbf{.081} (.032) &  .070 (.032) & .081 (.031) & .010 (.0022) \\ [0.5ex] \hline
WSe$_{2}$ & 300K & \textbf{.260} (.054) & .240 (.052) &  .190 (.053) & .160 (.053) & .022 (.014) \\ [0.5ex]
          & 150K & \textbf{.141} (.030) & .140 (.031) &  .081 (.031) & .070 (.031) & .010 (.004) \\ [0.5ex]
	  & 77K & .071 (.031) & \textbf{.082} (.031) &  .050 (.031) & .043 (.022) &  .011 (0.0021) \\ [0.5ex]
\hline 
\end{tabular}
\caption{Peak n-type (p-type) power factor of 1L, 2L, 3L, 4L and bulk MoS$_{2}$, MoSe$_{2}$,
WS$_{2}$ and WSe$_{2}$ at 300K, 150K and 77K.  The maximum power factor for each material at a given
temperature is in bold.}  
\label{tab:PF_max}
\end{table*}
\end{flushleft}

\begin{flushleft}
\vspace{0.2in}
\begin{table*}
\begin{tabular}{c c p{2.2cm} p{2.2cm} p{2.2cm} p{2.2cm} c}
\hline\hline
& Temperature & \multicolumn{1}{c}{1L} &  \multicolumn{1}{c}{2L}  & \multicolumn{1}{c}{3L} &  \multicolumn{1}{c}{4L} & Bulk  \\[0.5ex]  \hline
& & \multicolumn{5}{c}{\bf{Maximum n-type (p-type) ZT}} \\[0.5ex] \hline
MoS$_{2}$ & 300K & 1.35 (.970) & 1.35 (1.15) &  \textbf{2.23} (.510) & 1.78 (.390) & .350 (.081) \\ [0.5ex]
	& 150K & .590 (.350) & .590 (.450) &  \textbf{1.03} (.220) & .660 (.160) & .110 (.034) \\ [0.5ex]
	& 77K & .240 (.140) & .240 (.190) &  \textbf{.420} (.093) & .210 (.062) & .031 (.012)  \\[0.5ex] \hline
MoSe$_{2}$ & 300K & 1.39 (.800) & \textbf{2.39} (.810) &  1.66 (.810) & 1.65 (.810) & .290 (.081) \\ [0.5ex]
          & 150K & .450 (.310) & \textbf{1.06} (.320) &  .610 (.320) & .570 (.320) & .100 (.033) \\ [0.5ex]
	  & 77K & .130 (.120) & \textbf{.410} (.120)  &  .220 (.120) & .170 (.120) & .030 (.014) \\ [0.5ex] \hline
WS$_{2}$  & 300K & 1.52 (.70) & 1.98 (.720) &  \textbf{2.03} (.760) & 1.85 (.760) & .280 (.082) \\ [0.5ex]
          & 150K & .411 (.280) & .613 (.280) &  \textbf{.770} (.280) & .721 (.350) & .104 (.030) \\ [0.5ex]
	  & 77K & .120 (.110) & .181 (.113) &  .211 (.113) & \textbf{.271} (.113) & .034 (.012) \\ [0.5ex] \hline
WSe$_{2}$ & 300K & 1.88 (.620) & \textbf{1.92} (.620) &  1.44 (.620) & 1.13 (.620) & .260 (.120) \\ [0.5ex]
          & 150K & .590 (.220) & \textbf{.750} (.220) &  .490 (.220) & .380 (.220) & .091 (.032) \\ [0.5ex]
	  & 77K & .180 (.100) & \textbf{.270} (.100) &  .170 (.100) & .130 (.100) & .031 (.014) \\ [0.5ex]
\hline 
\end{tabular}
\caption{ Peak n-type (p-type) thermoelectric figure of merit, ZT, of 1L, 2L, 3L, 4L and 
bulk MoS$_{2}$, MoSe$_{2}$,WS$_{2}$ and WSe$_{2}$ at 300K, 150K and 77K.  
The maximum ZT for each material at a given temperature is in bold. }
\label{tab:ZT_max}
\end{table*}
\end{flushleft}

%

\end{document}